# Modelos Empíricos de Pós-Dupla Seleção por LASSO: Discussões para Estudos do Transporte Aéreo

Alessandro V. M. Oliveira

Instituto Tecnológico de Aeronáutica

Instituto Tecnológico de Aeronáutica. Praça Marechal Eduardo Gomes, 50. 12.280-250 - São José dos Campos, SP - Brasil.
E-mail address: alessandro@ita.br.

***Resumo*: O presente trabalho apresenta e discute formas de estimação por regressão regularizada e seleção de modelos com uso do método LASSO – Least Absolute Shrinkage and Selection Operator. O LASSO é reconhecido como um dos principais métodos de aprendizado supervisionado aplicados à econometria de alta dimensionalidade, permitindo trabalhar com grandes volumes de dados e múltiplos controles correlacionados. São abordadas questões conceituais relacionadas às consequências da alta dimensionalidade na econometria moderna e ao princípio da esparsidade, que fundamenta procedimentos de regularização. O estudo examina os principais modelos de pós-dupla seleção e pós-regularização, incluindo variações aplicadas a modelos de variáveis instrumentais. Também é apresentada uma breve descrição do pacote de rotinas lassopack, suas sintaxes e exemplos de modelos HD, HDS (High-Dimension Sparse) e IV-HDS, com combinações envolvendo estimadores de efeitos fixos. Por fim, discute-se o potencial de aplicação da abordagem em pesquisas voltadas ao transporte aéreo, com destaque para um estudo empírico sobre eficiência operacional de companhias aéreas e consumo de combustível de aeronaves.**

*Palavras-chave*: aprendizado de máquina, métodos de regularização, econometria de alta dimensionalidade, pós-lasso, variáveis instrumentais.

## I. Introdução

A popularização da internet e o avanço dos meios digitais na sociedade proporcionou o advento da proliferação dos chamados dados em grande volume – conhecidos como "*Big Data*". Nesse contexto, não apenas o número de observações contidas em diversas bases de dados aumenta consideravelmente a cada ano, mês, dia, hora e minuto, como também surgem mais e mais formas de se mensurar fenômenos – o que se traduz em conjuntos cada vez maiores de fatores explicativos passíveis de consideração nas análises quantitativas. Isso deu luz à chamada Estatística de "Alta Dimensão" – *High-Dimensional Statistics*, vide Giraud (2014), e Bühlmann & de Geer (2011). Em estatística, o termo "dimensionalidade" está atrelado ao número de atributos existente nos dados. Assim, dados, ou modelos, em alta dimensão são aqueles em que o número de variáveis é consideravelmente maior do que o utilizado na análise multivariada tradicional, gerando desafios importantes em termos computacionais e de interpretação de modelos. Sirimongkolkasem & Drikvandi (2019) e Liu, Xu & Li (2020) descrevem que a existência de dados de alta dimensão em campos como tecnologia da informação, astronomia, neurociência e bioinformática, em aplicações em genômica, análise de dados de ressonância magnética funcional, análise de dados de saúde em grande escala, análise de texto e imagem, dentre diversas outras, forçou o desenvolvimento de novos métodos de análise estatística.

Um dos métodos desenvolvidos para lidar com a dimensionalidade dos dados é o LASSO. O acrônimo LASSO (também referenciado como "Lasso" e "lasso") é utilizado em estatística para designar "*Least Absolute Shrinkage and Selection Operator*", ou, traduzindo literalmente, "Operador de Encolhimento Absoluto Mínimo e Seleção". Trata-se de um termo que deixa apenas implícito o que o modelo realmente faz, sendo que o mais apropriado seria denominá-lo de "Operador de Seleção e Encolhimento da Soma dos Valores Absolutos ao Mínimo". Trata-se de um método de análise de regressão que possibilita, simultaneamente à tradicional estimação dos parâmetros de mínimos quadrados, realizar:

1. o "encolhimento" de coeficientes, ou seja, a redução em módulo dos valores das estimativas, também conhecido como "regularização" e;
2. a seleção de variáveis, ou seja, a redução da dimensão do modelo, mantendo-se apenas um subconjunto do rol inicial de regressores.

Essas duas tarefas realizadas pelo LASSO - regularização e seleção -, são feitas simultaneamente no mesmo procedimento de otimização executado para se obter os coeficientes estimados[1]. Assim, ao encontrar a solução para os coeficientes que atende à restrição de soma dos valores absolutos dos mesmos, o LASSO ao mesmo tempo obtém coeficientes com valores menores em módulo, e obtém valores zerados para um subconjunto deles. Dessa forma, pode-se classificar o LASSO como um tipo de estimador de Mínimos Quadrados Restritos, cuja restrição força os coeficientes a tenderem a zero. Nesse sentido, trata-se de um procedimento de inferência estatística conservador, ao reduzir as chances de falsos positivos nas análises.

O LASSO tem origem em trabalhos pioneiros de Santosa e Symes (1986) e Frank e Friedman (1993), mas se tornou muito utilizado a partir do clássico estudo do professor da Universidade de Stanford Robert Tibshirani (1996), "*Regression shrinkage and selection via the lasso*", publicado no periódico britânico *Journal of the Royal Statistical Society: Series B (Methodological)*. No livro de Hastie, Tibshirani & Wainwright (2015), escrito quase vinte anos após a publicação do artigo, os autores refletem: "*O laço* [em português, com ç] *é uma corda longa com uma volta em uma das pontas, usada para pegar cavalos e gado. Em sentido figurado, o método "laça" os coeficientes do modelo*" (p. 8). "Laçar", nesse sentido, é selecionar apenas algumas variáveis dentre as disponíveis. Assim, trata-se de um método em que o pesquisador, de certa forma, automatiza a seleção das variáveis em sua especificação de modelo – em vez de proceder com a seleção manual, conhecida como seleção de subconjunto ("*subset selection*").

O LASSO é reconhecido universalmente como um dos métodos utilizados na área de aprendizado de máquinas (*machine learning*). Diz-se que o LASSO é uma das formas de de aprendizado supervisionado, ou seja, aquele em que um algoritmo trabalha sobre uma variável pré-definida de resposta que está contida nos dados – a variável dependente – e um objetivo específico – prever a variável dependente a partir de uma lista de variáveis independentes. Diz-se que os dados são "anotados" com as respostas, ou "rotulados" com as classes. De fato, para se proceder com uma estimação por LASSO, há que se primeiro especificar o regressando e todos os candidatos a regressores do modelo. Assumindo-se esparsidade, é possível, como veremos, interpretar que o modelo reduzido é mais eficaz em representar a realidade e fazer previsões. A abordagem de aprendizado supervisionado se distingue, por exemplo, do chamado "aprendizado não-supervisionado", em que as respostas ou rótulos nos dados não são previamente conhecidos ou disponíveis, e tudo o que se busca encontrar são padrões nos dados, ou seja, uma representação que produza alguma informação sintética dos mesmos, como por exemplo, uma em uma análise de agrupamentos (*clusters*).

## II. Seleção de modelos e o Princípio da Parcimônia

Segundo Burnham & Anderson (2004), uma frase famosa atribuída a Albert Einstein diz: "*Tudo deve ser feito o mais simples possível, mas não mais simples.*" (p. 30, livre tradução). Os autores argumentam que o sucesso na análise estatística de dados reais, e na inferência resultante dessa análise, muitas vezes depende da escolha do modelo de melhor aproximação aos dados. Modelos parcimoniosos deveriam ser perseguidos pelos pesquisadores, de forma a obter uma aproximação precisa da informação estrutural nos dados disponíveis, e que isso não deveria ser visto como uma busca pelo "modelo verdadeiro". Akaike (1974) sugere que a modelagem e a seleção de modelos estão essencialmente relacionadas à "*arte da aproximação*", que é "*um elemento básico da atividade intelectual humana*" (Akaike 1974, p. 716).

Perguntam Burnham & Anderson (2004) "*Se o ajuste é melhorado por um modelo com mais parâmetros, onde devemos parar?*". Em outras palavras, existe uma regra de parada na inserção de mais variáveis em um modelo? Trata-se de uma questão clássica, constatada na medida em que se descobriu que estatísticas de ajuste como o $R^2$ em geral aumentam com o número de variáveis adicionadas aos modelos – o que gera uma consequência não desejada de incentivo à inserção de complexidade desnecessária pelos pesquisadores. O chamado "Princípio da Parcimônia" advém de Box e Jenkins (1970), que sugere que a tarefa de especificação deveria levar a um modelo com "(...) *o menor número possível de parâmetros para representação adequada dos dados.*" (p. 17). Segundo Burnham & Anderson (2004), os estatísticos enxergar o princípio da parcimônia como um balanceamento (*trade-off*) entre viés e variância. Viés seria o quão distante nosso modelo prevê os dados utilizados na estimação – dados amostrais, *in-sample predictions* – quando obtidas muitas amostras. Variância seria o quanto nosso modelo erra nas previsões, na medida em que usamos dados fora da amostra (*out-of-sample predictions*). A Figura 1 ilustra esse *trade-off*, problema esse que, de certa forma, está presente em todos os métodos de seleção de modelo. Dizem os autores, que, em geral, o viés diminui (curva vermelha) e a variância das previsões fora da amostra aumenta (curva azul), na medida em que a complexidade do modelo, medida por sua dimensão (número de parâmetros $p$) aumenta. Frequentemente, podemos usar o número de parâmetros em um modelo como uma medida do grau de estrutura exigido dos dados. Modelos parcimoniosos alcançam um balanço adequado entre viés e variância, localizando um modelo de complexidade ideal, ótima, ditado pelo ponto de mínimo da curva de erro total, em preto. Um nível de complexidades abaixo do ideal gera o subajuste (*underfitting*), enquanto que um nível acima gera o sobreajuste (*overfitting*).

---

[1] Note que o LASSO requer um procedimento prévio de padronização das variáveis, ou seja, transformação de maneira a obter média igual a zero e desvio padrão igual a 1.

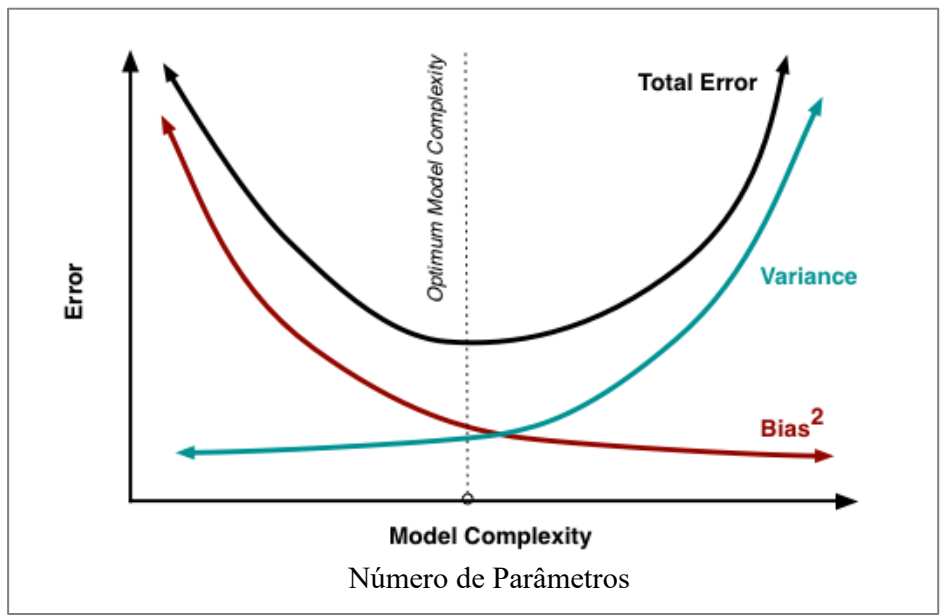

**Figura 1 – Tradeoff entre Viés e Variância**
**(Burnham & Anderson, 2004, p. 31 e**
**ethen8181.github.io/machine-learning/regularization/regularization.html)**

## III. A Maldição da Dimensionalidade

Segundo Giraud (2014), conjuntos de dados estatísticos estão crescendo rapidamente em muitos domínios devido ao desenvolvimento de avanços tecnológicos que ajudam a coletar dados com um grande número de variáveis para melhor compreender um determinado fenômeno de interesse. Uma das principais características dos dados modernos é que eles geralmente são obtidos a partir de coletas simultâneas de milhares a milhões de informações em cada objeto ou indivíduo. Esses dados são considerados de "alta dimensão", e que também induzem o uso de modelos de "alta dimensão".

Temos um problema de estimação em alta dimensão quando o número de preditores (dimensões) $p$ é grande - Ahrens, Hansen & Schaffer (2020). Em casos extremos, e cada vez mais observados em muitas áreas do conhecimento, o pesquisador se defronta com dados altamente dimensionais em que $p$ é de magnitude comparável, ou até mesmo muito maior que o tamanho da amostra $n$ – Liu, Xu & Li (2020) –, caso este em que a estimação por Mínimos Quadrados Ordinários (OLS) é inviabilizada pela simples falta de graus de liberdade no procedimento para se estimar o modelo completo. Temos assim, pelo menos duas formas de enquadrar a alta dimensão dos dados: uma absoluta ($p$ grande) e outra relativa ($p$ grande quando comparado a $n$, em geral maior que este). Há também o caso extremo em que a maldição da dimensionalidade é aguda, com o número de variáveis explicativas é muito maior do que o número de pontos de dados ($p \gg n$, caso denominado de "*short, fat data problem*", vide Clarke & Chu, 2014). Denotaremos os modelos (ou dados) de alta dimensão com a sigla "HD", a partir de sua classificação absoluta, ou seja, independente de estarmos ou não em uma situação de falta de graus de liberdade suficientes para a estimação de parâmetros.

O matemático norte-americano Richard E. Bellman cunhou a expressão "Maldição da Dimensionalidade" ao descrever o problema causado pelo aumento exponencial de volume associado à adição de dimensões extras a um espaço matemático, com implicações sobre o tempo computacional de métodos de solução numérica na área de programação dinâmica. Assim, na medida em que o número de dimensões aumenta, a quantidade de dados que um modelo necessita para propiciar generalizações dos fenômenos aumenta exponencialmente. E, adicionalmente, o valor adicionado por uma dimensão a mais inserida na resolução de um problema pode ser muito menor em comparação com a sobrecarga que ele adiciona ao modelo ou algoritmo que o irá analisar. A maldição da dimensionalidade em geral se refere a vários fenômenos que possivelmente surgem ao se analisar, e mesmo organizar, dados em espaços de alta dimensão. Sirimongkolkasem & Drikvandi (2019) argumentam que fenômenos dimensionalmente amaldiçoados ocorrem em domínios não apenas da análise numérica, mas também na amostragem, combinatória, aprendizado de máquina, mineração de dados e bancos de dados. Dizem os autores que o tema comum desse tipo de problema é que, para obter um resultado estatisticamente significante, sólido e confiável, a quantidade de dados necessária para suportar o resultado geralmente cresce exponencialmente com a dimensionalidade.

### III.1. EXEMPLO I: MICRODADOS DE QUESTIONÁRIOS JUNTO A PASSAGEIROS

Para ilustrar um pouco a questão da dimensionalidade, suponha que tenhamos dados obtidos a partir de uma coleta de questionários (*survey*) junto a passageiros em um aeroporto. Trata-se de um exemplo baseado em questionários reais, aplicados nos últimos anos no Brasil – *surveys* da Fundação Instituto de Pesquisas Econômicas, FIPE, em 2009, e da Empresa de Planejamento e Logística S.A. EPL, em 2014. Em nosso exemplo fictício, foram obtidos 500 questionários em um dado dia, contendo as mais diversas informações socioeconômicas e comportamentais dos passageiros, com a finalidade de se estudar o seu comportamento de consumo de viagens aéreas. Dentre os atributos coletados, temos os listados a seguir. Para alguns deles, imagine que a informação é obtida complementando-se com uma consulta a fontes secundárias de dados – por exemplo, o passageiro em geral não saberia responder qual o tipo de aeronave que irá embarcar, mas essa informação pode ser obtida consultando-se posteriormente a aeronave utilizada em seu voo.

**Dados da viagem**
- Aeroporto da pesquisa
- Segmento de Viagem
- Tipo de voo
- Par de aeroportos do voo Perna de voo
- Hora do voo
- Tipo da aeronave a ser utilizada

**Dados de perfil do passageiro**
- Par OD real
- Nacionalidade
- Gênero
- Idade
- Renda
- Número de dependentes
- Ocupação
- Escolaridade
- Quantas viagens por ano faz por avião?
- Quantas viagens não urbanas/rotineiras faz por ano faz por outros modos de transporte?
- Número de viagens domésticas nos últimos 12 meses
- Número de viagens internacionais nos últimos 12 meses
- Realizou viagens por outro modo de transporte neste trecho de viagem?

**Dados econômicos da compra efetuada**
- Companhia aérea/par de aeroportos/dia/hora escolhidos
- Número de conexões de voo
- Número de escalas de voo
- Companhias aéreas/par de aeroportos/dia/hora ofertados nesse mercado
- Pesquisou passagem aérea em outros dias ou outros aeroportos?

- Voo codeshare ou de parceria?
- Preço da passagem
- Tem programa de milhagem nessa companhia aérea?
- Tem programa de milhagem em outras companhias aéreas?
- Comprou usando milhas?
- Quantas milhas gastou?
- Despachou bagagem?
- Valor da taxa de despacho de bagagem
- Valor da taxa de remarcação
- Outras taxas pagas
- Dias de antecedência da compra da passagem
- Quem pagou a passagem aérea?

**Dados comportamentais**

- Tempo de trajeto terrestre até o aeroporto de origem
- Tempo de trajeto terrestre a partir do aeroporto de destino
- Tempo de permanência no aeroporto
- Tipo de *check-in*
- Modo de acesso ao aeroporto de origem
- Modo de acesso a partir do aeroporto de destino
- Motivo da viagem
- Comprou algo no aeroporto?
- Total de gastos no aeroporto

Ao observar o conjunto de dados coletados dessa nossa pesquisa fictícia junto a cada um dos 500 entrevistados, podemos chegar rapidamente à conclusão de que muito mais de 500 variáveis podem ser construídas a partir dessas informações. Por exemplo, só de categorias de dados de perfil (gênero, ocupação, etc.), poderíamos chegar a umas dezenas de *dummies*, aplicando o chamado "*one-hot encoding*" – ou seja, a transformação de cada categoria em variáveis binárias distintas (colunas da matriz de dados), contendo o valor "1" caso o indivíduo pertença àquela categoria, e "0", caso não pertença. Suponha que, apenas considerando as variáveis de perfil, obtivéssemos 50 *dummies*. Já as categorias temporais, como dias de antecedência de compra, e hora do voo, poderiam ser representadas por, digamos, 120 *dummies* de dias de antecedência e 24 *dummies* de hora de partida do voo – ou, a depender da modelagem, 24 × 7 = 168 dummies, para contar a semana cheia. Se considerarmos o *one-hot encoding* para cada par de aeroportos ou cidades, poderíamos ter centenas deles, a depender da amostragem. Digamos que fossem 150 os pares de aeroporto de origem - aeroporto de destino do passageiro, 250 pares de cidade efetiva (real) de origem - cidade efetiva (real) de destino do passageiro e mais uns 50 pares de aeroporto de origem-destino do voo. Temos que, somente com essas *dummies* geradas, chegaríamos a 788 dimensões, o que inviabilizaria a estimação por Mínimos Quadrados Ordinários. Temos configurado um problema de alta dimensionalidade.

Uma nota importante diz respeito à "armadilha da variável *dummy*" (*dummy variable trap*). É classicamente conhecido que o uso de *dummies* para todas as categorias de uma variável qualitativa gera problemas de inversibilidade da matriz no estimador de Mínimos Quadrados. Por exemplo, em uma regressão com intercepto, se incluirmos uma *dummy* de sexo feminino e outra de sexo masculino, representativos da variável "gênero" com duas categorias coletadas, teremos a impossibilidade da obtenção das estimativas por conta da colinearidade perfeita entre essas variáveis. Esse certamente também é o caso do exemplo dado acima, da *survey* fictícia junto a passageiros. Para cada variável categórica, uma categoria terá que ser deixada como "caso de referência" ou "caso base", ou seja, não será reportado e forçará que cada um dos coeficientes das demais categorias sejam interpretados relativamente a ele. Esse é um problema distinto do problema da alta dimensionalidade reportado acima, mas igualmente gera a impossibilidade da estimação por mínimos quadrados. Entretanto, como veremos, alguns estimadores de regularização, como o LASSO, lidam normalmente com variáveis perfeitamente colineares – na verdade, dropam uma (ou mais) delas automaticamente –, não sendo necessário informar uma categoria de referência quando da inicialização da rotina.

### III.2. Exemplo II: dados agregados de demanda por viagens

Ainda para ilustrar a questão da dimensionalidade em dados do transporte aéreo, suponha algumas situações de agregação de dados em um estudo de demanda por viagens aéreas no Brasil. Primeiramente, considere dados agregados ao nível nacional. Essa é a antítese do caso dos microdados de passageiros, vistos no Exemplo I. Um estudo estudo econométrico com esse tipo de dados agregados poderia conter o número de passageiros voados (ou o total de passageiros-quilômetros transportados) como regressando, rodado contra fatores como a renda média brasileira, o preço (ou *yield*) médio das empresas aéreas, e talvez um conjunto de alguns outros controles, como variáveis *dummies* representativa de eventos exógenos, dentre outros. Em suma, teríamos um modelo com poucos dados, mas também com poucos regressores.

Continuando no exemplo da demanda por viagens aéreas, suponha agora que temos dados agregados ao nível do par de cidades. Um estudo de demanda ao nível do par de cidades iria requerer a observabilidade, no mínimo, da renda média na origem e destino dos mercados servidos – ou alguma média desses valores –, e o preço médio das passagens aéreas nesses mercados, mas também a distância entre origem e destino. Diversos outros fatores específicos das rotas poderiam ser considerados, como outras características relevantes das cidades envolvidas – desemprego, desigualdade de renda, número de frequências de voo, congestionamento dos aeroportos envolvidos, etc. – o que poderia levar o estudo a ter um conjunto mais amplo de regressores. Se fosse um painel de dados – pares de cidades observadas ao longo do tempo –, *dummies* (ou efeitos fixos) de pares de cidade e de períodos poderiam ser utilizados, dentre outras possibilidades. Se o nível de agregação fosse o par de aeroportos, em vez do par de cidades, seria fundamental ter o controle de variáveis indicativas da concorrência aeroportuária, ou seja, do quanto as condições de preço e outros atributos de um par de aeroportos poderia influenciar a demanda de outro par de aeroportos dentro de uma mesma região de múltiplos aeroportos. Esses controles levariam a um aumento adicional na sobrecarga da dimensionalidade dos modelos.

Por fim, para fins de comparação com o Exemplo I, imagine a nossa *survey* fictícia, ou seja, que tivéssemos o luxo de poder de coletar microdados de passageiros. A pesquisa junto a passageiros em aeroportos aplica um questionário diversas de questões do perfil, atitude e hábitos de consumos dos entrevistados, que proporciona um verdadeiro "*zoom in*" da observabilidade, uma análise microscópica da realidade de cada passageiro, em vez de uma agregação onde toda essa informação extremamente relevante é perdida. Para realizar um estudo de demanda nesse nível de desagregação, teríamos que desenvolver um problema de escolha discreta – *probit*, *logit*, e suas vertentes mais modernas –, onde todas as alternativas de consumo à disposição do passageiro deveriam ser controladas – outras ofertas da mesma companhia aérea naquele par de aeroportos e em pares de aeroportos adjacentes, e outras ofertas das demais companhias aéreas naquele mercado de viagens. Assim, adicionalmente às características socioeconômicas e comportamentais, o banco de dados da pesquisa deveria conter todas as informações do ambiente competitivo específico de cada situação de consumo.

Podemos observar, com a ilustração do problema de estimação da demanda por viagens, que, alterando o nível de agregação dos dados, a dimensionalidade necessariamente irá aumentar – de um banco de dados com poucas variáveis, pode-se chegar a estudos com centenas e mesmo milhares delas – o que requer um conjunto de dados mais amplo. O incremento da dimensionalidade traz consigo o aumento da complexidade inerente ao fenômeno, quando observado em alto nível de desagregação, o individual – um problema estudado pela microeconometria. Em nosso exemplo, o problema da dimensionalidade não chega a inviabilizar a estimação tradicional, dado que o número de dimensões requeridas cresce menos rapidamente do que o número de dados obtidos para estudo. Ou seja, com um aumento das amostras, é possível prosseguir com o desenvolvimento de modelos econométricos. Em algumas áreas do conhecimento, entretanto, essa problemática pode, efetivamente, se transformar em uma maldição da dimensionalidade, com a necessidade de dados crescendo exponencialmente com o número de dimensões adicionais inseridas. Por outro lado, a existência de muitas dimensões em um modelo econométrico em qualquer área pode gerar problemas importantes de estimação, como multicolinearidade, interpretabilidade e sobreajuste.

# IV. Consequências da Alta Dimensionalidade

Banks (2019) apresenta três descrições aproximadamente equivalentes da "maldição da dimensionalidade": 1. para o tamanho amostral $n$ fixo, na medida em que $p$ aumenta, os dados se tornam esparsos; 2. na medida em que $p$ aumenta, o número de modelos possíveis explode; 3. para um $p$ alto, a maioria das bases de dados se torna multicolinear.

Comecemos com o último ponto destacado por Banks (2019). Um dos efeitos prováveis da estimação de modelos HD – e mesmo em situações em que o número de parâmetros não é tão elevado em relação ao tamanho da amostra –, é a geração, ou agravamento, do problema de multicolinearidade. Multicolinearidade é a situação em que duas ou mais variáveis preditoras em um modelo estatístico estão linearmente relacionadas – Gujarati & Porter (2011) e Wooldridge (2016). Em um caso extremos de multicolinearidade, teríamos a chamada "colinearidade perfeita" entre regressores. Suponha o caso em uma regressão linear de $Y$ em $X_1$ e $X_2$, onde $\beta_0$ é o intercepto, $\beta_1$ e $\beta_2$ são parâmetros de inclinação, e $\varepsilon$ é o erro aleatório. Suponha uma colinearidade perfeita simples, em que $X_1 = X_2$. Nesse caso, poderíamos ter, equivalentemente:

$$Y = \beta_0 + \beta_1 X_1 + \beta_2 X_2 + \varepsilon, \qquad (1)$$

ou

$$Y = \beta_0 + (\beta_1 + \beta_2)X_1 + 0 X_2 + \varepsilon, \qquad (2)$$

ou

$$Y = \beta_0 + 0 X_1 + (\beta_1 + \beta_2)X_2 + \varepsilon, \qquad (3)$$

ou

$$Y = \beta_0 + (\beta_1 + k)X_1 + (\beta_2 - k)X_2 + \varepsilon. \qquad (4)$$

Ou seja, qualquer soma de $\beta_1$ e $\beta_2$ poderia ser "encontrada" pelo estimador em uma solução, inclusive com termos compensados entre parâmetros usando uma constante $k$. Nesse último caso, poderia ocorrer de haver troca de sinal, sendo esse efeito compensado no outro coeficiente, que ficaria "inflado" em termos absolutos. Essa possibilidade gera o que os manuais de econometria se referem a uma "instabilidade" dos parâmetros devido à multicolinearidade. Pode haver, assim, a estimação de coeficientes com valores absolutos altos, sendo que o valor de um compensa o valor do outro. Em nosso caso acima de colinearidade perfeita, não será possível estimar os coeficientes por OLS. Entretanto, em um caso onde a colinearidade entre regressores é alta, mas não perfeita, fenômenos similares podem ocorrer, prejudicando a inferência estatística.

Assim, a vigência de multicolinearidade alta em uma amostra de dados pode resultar em um aumento da variância dos coeficientes de regressão, o que gera uma estimativa instável dos valores dos parâmetros. Nesse caso, a instabilidade dos parâmetros se manifesta em algumas situações: 1. os coeficientes às vezes parecem ser coeficientes insignificantes, mas na realidade são relações significativas entre um coeficiente preditor com o coeficiente de resposta – <u>risco de falso negativo</u> por aumento da variância estimada do coeficiente, com identificação errônea de preditores relevantes; 2. os valores dos coeficientes dos regressores podem variar bastante de uma amostra para outra – quando se obtém uma nova amostra, ou mesmo quando se efetua pequenas alterações nos dados, como o experimento de descarte de algumas observações; 3. a remoção de um termo afeta bastante o valor absoluto dos demais regressores.

Em suma, em um modelo econométrico, a multicolinearidade severa torna-se um problema porque ela pode aumentar a variância das estimativas dos coeficientes e tornar as estimativas muito sensíveis a pequenas alterações no modelo. O resultado pode ser que as estimativas dos coeficientes serão instáveis e difíceis de interpretar. Em alguns casos, como vimos, dois ou mais coeficientes são estimados com valores altos com sinais opostos que se compensam e se um deles for descartado, um grande impacto é ocasionado no(s) outros(s). A multicolinearidade mina o poder estatístico da análise, podendo fazer com que os coeficientes mudem de sinal, tornando mais difícil especificar o modelo correto e menos confiável o modelo final escolhido.

O uso de dados HD pode não apenas levar à identificação errônea de preditores relevantes dentro de uma regressão por conta da multicolinearidade, mas também pode impactar a capacidade do modelo de extrapolar além do intervalo da amostra com a qual foi construído – isto é, aumenta o risco do chamado "sobreajuste" (*overfitting*). Em estatística, o sobreajuste ocorre quando um modelo consegue efetuar previsões que são muito próximas aos valores observados de um determinado conjunto de dados utilizados para a estimação do modelo – ou seja, produz excelentes previsões *dentro da amostra* –, mas que perde desempenho em previsões utilizando dados adicionais ou observações futuras – ou seja, produz previsões *fora da amostra* consideradas piores ou abaixo do esperado. Em geral um modelo que padece de sobreajuste contém mais parâmetros do que pode ser justificado pelos dados e, assim, temos que esse tipo de problemática é mais comum de emergir com dados HD. Segundo Burnham & Anderson (2004), no cerne do problema do sobreajuste está o uso de muitas variáveis irrelevantes, que acabam captando, sem o pesquisador saber, um efeito espúrio contido nos resíduos, apresentando esse efeito nos resultados como se essa variação representasse a estrutura do modelo subjacente aos dados. Na existência de dados HD, é relativamente fácil para o pesquisador incorrer no problema do sobreajuste. Segundo Tibshirani (1996), existem duas razões pelas quais o analista de dados geralmente não fica satisfeito com as estimativas OLS. A primeira é a precisão da previsão: as estimativas OLS geralmente têm baixo viés, mas grande variância – ou seja, erram muito facilmente as previsões fora da amostra. O autor explica que a precisão da previsão pode, às vezes, ser melhorada reduzindo ou definindo como 0 alguns coeficientes – o procedimento do LASSO que ele propõe no artigo. "*Ao fazer isso, sacrificamos o modelo com um pouco de viés para reduzir a variância dos valores previstos e, portanto, podemos melhorar a precisão geral da previsão*" - Tibshirani (1996, p. 267, livre tradução).

A segunda razão dos analistas não ficarem satisfeitos com o OLS, segundo Tibshirani (1996) é a interpretação dos modelos. "*Com um grande número de preditores, muitas vezes gostaríamos de determinar um subconjunto menor que exibe* [apenas] *os efeitos mais fortes*" (1996, p. 267, livre tradução). De fato, como vimos ao discutir multicolinearidade, a especificação de muitas variáveis em um modelo pode aumentar o risco que algumas delas tenham correlação linear entre si, o que torna mais complexo de interpretar o efeito isolado – chamado de "efeito *ceteris paribus*" – de cada uma delas. Temos configurado mais um desafio para os pesquisadores que trabalham com dados e modelos de alta dimensão.

Para concluir, temos que o uso de dados HD configura um conjunto de armadilhas para os pesquisadores da área. Em especial, a redundância informacional no uso de variáveis regressoras, que dá origem a problemas de multicolinearidade, sobreajuste e comprometem a interpretabilidade dos modelos. Idealmente, os modelos de regressão deveriam se valer de variáveis preditoras que correlacionam altamente com o regressando, mas se correlacionam no muito pouco entre si. Isso incrementa a qualidade da interpretação *ceteris paribus* dos coeficientes. Um modelo assim ideal é por vezes chamado de modelo com "baixo ruído" e será capaz de fazer previsões confiáveis em várias amostras de conjuntos de variáveis retiradas da mesma população estatística.

O bioestatístico da Johns Hopkins Bloomberg School of Health Jeff Leek descreve o outro lado da moeda da maldição da dimensionalidade, conhecido como "bênção da dimensionalidade" (*blessing of dimensionality*): "*Basicamente, uma vez que um número cada vez maior de medições é feito nas mesmas observações, há uma estrutura inerente a essas observações. Se você tirar vantagem dessa estrutura, conforme a dimensionalidade do seu problema aumenta, você obtém melhores estimativas da estrutura em seus dados de alta dimensão - uma bela bênção!*" (Leek, 2015). Ilustra o ponto partir de uma ilustração: "*Como exemplo, suponha que façamos medições em 10 pessoas. Começamos fazendo uma medida (pressão arterial), depois outra (altura), depois outra (cor do cabelo) e continuamos indo e indo até termos um milhão de medidas nessas mesmas 10 pessoas. A bênção ocorre porque as medições dessas 10 pessoas estarão todas relacionadas entre si. Se 5 das pessoas forem mulheres e 5 ou homens, então qualquer medida que tenha uma relação com sexo estará altamente correlacionada com qualquer outra medida que tenha uma relação com sexo. Portanto, conhecendo um pequeno pedaço de informação, você pode aprender muito sobre muitas das diferentes medições*" (Leek, 2015).

A bênção da dimensionalidade nos permite discorrer sobre as vantagens de se possuir dados HD: tendo em mãos um conjunto mais completo de controles para fenômenos observáveis e não observáveis. Sabendo da existência de "*nuisance variables*" ("variáveis de incômodo") – ou seja, fatores latentes subjacentes ao problema investigado, mas que são "indesejados" no sentido de não serem de interesse para o pesquisador em um determinado estudo, e que em que podem ser, ou em geral são, correlacionados com a(s) variável(is) independente(s) hipotética de interesse em seu estudo. Essas variáveis de incômodo podem, por exemplo, ser características não-observáveis dos indivíduos de um painel ou *cross-section* participantes em estudo, ou qualquer influência não intencional do pesquisador, em uma manipulação experimental (Salkind, 2010), ou dos indivíduos sob análise, ou mesmo de fatores outros não antecipados. Haig (1992) descreve que essas são

"variáveis terceiras" (pois interferem na relação entre X e Y) e que são denominadas de "*nuisance*" na medida em que são controladas estatisticamente de alguma forma no estudo, de maneira a permitir a inferência de causalidade das variáveis de interesse. Quando o controle estatístico de variáveis terceiras não é realizado, ou é efetuado de maneira insatisfatório, em geral encontra-se na literatura a denominação de "*confounding variable*" (variável de confusão de efeitos) para essas variáveis – variáveis latentes que, ao falhar em controlar, prejudicam a validade interna de um experimento e/ou as análises de relação de causalidade e impossibilitando inferências *ceteris paribus* (efeitos isolados de uma variável). De posse de dados HD, as possibilidades de controles de efeitos não observáveis aumenta consideravelmente, por meio de *dummies* de *clusters* de indivíduos semelhantes, efeitos fixos de indivíduos em uma estrutura de painel de dados, efeitos temporais, ou outras formas de controle que são específicas de cada caso. O número de controles utilizados pode ser elevado, a depender da complexidade do fenômeno, mas pelo fato de haver a possibilidade de serem incluídos no estudo e permitirem o controle das variáveis terceiras, melhoram qualitativamente as inferências estatísticas – o que é uma das vertentes da bênção da dimensionalidade.

Um ponto adicional que reforça a relevância dos controles HD é a possibilidade de traçar um paralelo entre o tratamento de fatores não observáveis e o propósito do teste RESET de Ramsey, tradicionalmente utilizado em econometria. Enquanto o primeiro busca incorporar diretamente múltiplos controles que representem heterogeneidades observáveis e não observáveis, o segundo avalia indiretamente se tais fatores latentes ou não linearidades permanecem não controlados no modelo. O teste RESET, ao introduzir potências dos valores ajustados para verificar se ainda há explicação adicional da variável dependente, sinaliza possíveis omissões de variáveis ou formas funcionais inadequadas, muitas vezes associadas a efeitos de variáveis não observadas. Já a ampliação dimensional com efeitos fixos, dummies de grupos ou controles temporais fornece mecanismos empíricos para capturar essas complexidades estruturais e reduzir o viés por variáveis omitidas. Em síntese, o teste RESET aponta a insuficiência de controle sobre não linearidades e fatores latentes, ao passo que os dados HD representam a via prática para superá-la.

# V. Princípio da Esparsidade

Uma forma de lidar com dados HD seria a chamada "seleção de subconjunto" (*subset selection*), conforme discute Tibshirani (1996). A seleção de subconjunto refere-se à tarefa de encontrar um pequeno subconjunto das variáveis independentes disponíveis a partir de buscas exaustivas e visando atender a algum critério pré-selecionado. Uma das formas de seleção de subconjunto é a *stepwise regression*, tipo de regressão "passo-a-passo" muito criticada, seja em sua forma automática – usando algoritmo – ou em sua forma manual – experimentação de modelos feita corriqueiramente pelos pesquisadores – no senso comum, o chamado "tira e põe" de variáveis.

Modernamente, já é bastante conhecida a impropriedade de comportamentos de especificação *ad-hoc* de modelos denominados como "*data fishing*" ou "*p-hacking*" – quando se faz uso indevido da análise de dados para encontrar padrões em dados que podem ser apresentados como estatisticamente significativos, aumentando e subestimando dramaticamente o risco de falsos positivos e também relatando apenas aqueles que apresentam resultados significativos. Em relação a procedimentos automáticos, conforme observa Smith (2018), em um artigo com trocadilhos denominado "*Step away from stepwise*" ("Afaste-se do passo-a-passo"), um problema fundamental da regressão *stepwise* é que alguma variável explicativa real que tem efeitos causais na variável dependente pode não ser estatisticamente significativa, enquanto uma variável não observada e sem interesse direto na análise pode ser coincidentemente significativa. Como resultado, o modelo pode se ajustar bem aos dados dentro da amostra, mas se sair mal fora da amostra. Tibshirani (1996) destaca que a seleção de subconjuntos pode até fornecer modelos interpretáveis, mas também pode ser extremamente variável porque é um processo discreto - os regressores são ou retidos ou eliminados do modelo. Diz o autor que "*Pequenas alterações nos dados podem resultar na seleção de modelos muito diferentes e isso pode reduzir a precisão da previsão*".

Um artifício comumente utilizado por esse tipo de abordagem que busca que lidar com dados HD é o princípio ou pressuposto de "esparsidade" (*sparsity*) dos modelos. Esparsidade é o equivalente da ideia do senso comum em que "menos é mais". Mais formalmente, pode-se dizer que esparsidade se refere ao fenômeno de que "*uma estrutura de dados subjacente pode ser explicada principalmente por poucos de muitos recursos*" – Sirimongkolkasem & Drikvandi (2019). Por exemplo, se temos um conjunto elevado de variáveis potencialmente explicativas de um determinado regressando, ao utilizar o pressuposto de esparsidade estamos assumindo que apenas um subconjunto reduzido desses regressores realmente pode ajudar na predição do modelo. De uma maneira mais estrita, poderíamos assumir que a esparsidade também implicaria em que apenas esse subconjunto reduzido de regressores de fato pertenceria ao processo gerador dos dados – ou se refere ao "modelo verdadeiro" ou à "realidade" –, ou seja, teria papel na determinação do fenômeno a ser explicado. Esse contexto é ainda mais desafiador especialmente sabendo que a situação de pesquisa mais comumente observada acontece quando modelo verdadeiro (subjacente aos dados) é tratado como desconhecido ao pesquisador. Entretanto, assumir uma interpretação tão estrita de esparsidade – isto é, atrelada ao processo gerador dos dados – nem sempre é necessário para fins práticos de estimação, dado que estamos interessados em incrementar a interpretabilidade dos modelos e reduzir a redundância informacional no uso das variáveis explicativas. Burnham & Anderson (2004) evitam o esse uso dos termos "*underfitted*" (subajustado) e "*overfitted*" (sobreajustado) em um sentido que supõem a existência de um modelo "verdadeiro" de baixa dimensão como um padrão. Em vez disso, dizem os autores (p. 32, livre tradução) "*reservamos os termos subajustado e sobreajustado para uso em relação a um "modelo de melhor aproximação"*". A discussão dos autores leva a uma preferência de modelos sobreajustados, e vai na seguinte linha de raciocínio. Por um lado, modelos sobreajustados são frequentemente livres de viés nos estimadores de parâmetros, mas têm variâncias desnecessariamente grandes, com baixa precisão dos estimadores em relação ao que poderia ter sido realizado com um modelo mais parcimonioso. Adicionalmente, efeitos espúrios de tratamento tendem a ser identificados, dado que variáveis espúrias podem ser incluídas em modelos sobreajustados. Por outro lado, modelos subajustados iriam ignorar alguma estrutura replicável importante nos dados – o que significa dizer conceitualmente replicável na maioria das outras amostras – e, portanto, geraria viés substancial, com variância amostral subestimada – ambos os fatores resultando em problemas na estimação dos intervalos de confiança. A tendência é que modelos subajustados (*underfitted*) sejam um problema mais sério na análise de dados e inferência do que modelos sobreajustados (*overfitted*) – Shibata (1989).

Segundo Banks (2019), a intuição por trás da abordagem do problema da maldição da dimensionalidade é "*a aposta no princípio da esparsidade, que diz que, em altas dimensões, é prudente prosseguir sob o pressuposto de que a maioria dos efeitos não é significante*". Diz o autor que esse princípio pode ser justificado de duas formas: 1. é, em geral, verdadeiro (por meio da Navalha de Occam – que reza que a explicação mais simples para uma resposta é a mais provável de ser a correta – ser empiricamente sólida); e 2. se o problema não for esparso, então não você não será capaz de fazer algo útil sobre ele de qualquer maneira. Ainda Banks (2019), mesmo que a esparsidade não seja estrita – ou seja, os efeitos que chamamos de esparsos não são de fato nulos –, pode ocorrer a situação em que um subconjunto reduzido deles seja capaz de explicar uma grande proporção da variabilidade nos dados.

Por meio da suposição de esparsidade, é possível avaliar que métodos de seleção como a estimação por LASSO podem ser considerados "Modelos Econométricos Esparsos de Alta Dimensão" (*High-Dimensional Sparse Econometric Models*) – Belloni & Chernozhukov (2011), Belloni, Chernozhukov & Hansen (2013). Isso porque o LASSO possibilita justamente atingir uma solução esparsa – diversos coeficientes zerados são obtidos na solução ótima do problema –, ou seja, um portfolio de regressores selecionados que é reduzido em relação ao portfolio original. Dessa forma, denotaremos esse modelo com a sigla "HDS", em contraposição aos modelos puramente "HD", onde nenhum procedimento de seleção para reduzir a dimensão foi aplicado. Ainda referente a terminologias, os autores em geral utilizam o conceito de "solução esparsa" permitida por um modelo. Por exemplo, Sirimongkolkasem & Drikvandi (2019, p. 740) discorrem que "(...) *o LASSO fornece uma solução esparsa (ou seja, o número de estimativas de parâmetros diferentes de zero é menor do que o tamanho da amostra n)*"

Sirimongkolkasem & Drikvandi (2019) atestam ainda que a regressão LASSO requer o pressuposto de esparsidade, ou seja, muitas das covariadas devem ser consideradas não relacionadas à variável de resposta – nossa interpretação estrita da esparsidade, conforme discutido acima. Dizem os autores que assim, o modelo se torna atraente aos pesquisadores na área de dados HD, dado ser capaz de identificar, dentre um grande número de preditores, um punhado deles que são as principais contribuições para algumas previsões desejadas – citam, como exemplo, os estudos de associação genômica ampla – GWAS, estudos que associam características dos indivíduos e conjunto de genes associados. "*Isso leva a modelos parcimoniosos a partir dos quais a variável selecionada pode ser examinada mais detalhadamente, além de reduzir significativamente os custos computacionais subsequentes nas previsões*" - Sirimongkolkasem & Drikvandi (2019, p. 740).

Ahrens, Hansen & Schaffer (2020) utilizam o seguinte índice de esparsidade $s$:

$$s := \sum_{j=1}^{p} \mathbf{1}\{\beta_j \neq 0\}, \tag{5}$$

onde $p$ é o número de preditores, $\beta_j$ é o $j$-ésimo parâmetro do modelo econométrico, $\mathbf{1}\{\beta_j \neq 0\}$ é uma função indicadora, sendo igual a 1 caso o respectivo parâmetro seja não nulo, e zero caso contrário. Trata-se do número de elementos não nulos do vetor de parâmetros, também conhecido como "pseudo-norma" ou $\ell_0$). Quanto menor $s$, mais esparso é o modelo, no sentido que um subconjunto mais reduzido de parâmetros é incluído[2]. Assim, a esparsidade se uma solução para um problema de estimação com dados em alta dimensão implica que a contagem de coeficientes não-nulos obtida $s$ satisfaça à seguinte propriedade:

$$\{s \ll p, s \ll N\}, \tag{6}$$

onde $N$ é o número de observações. Essa solução é denominada de "solução esparsa", ou "aproximação esparsa $\ell_0$". Os autores também definem o que chamam de "conjunto ativo" de coeficientes do modelo ($\Omega$), que abrange todos coeficientes não nulos:

$$\Omega := \{j \in \{1, \ldots, p\} : \beta_j \neq 0\}, \tag{7}$$

## VI. Regressão linear regularizada

Ao reduzir a dimensão do modelo por meio da seleção embutida no problema de otimização de mínimos quadrados, os métodos de regularização, como a regressão RIDGE e LASSO, permitem suavizar os problemas associados à multicolinearidade e à complexidade dos modelos econométricos HD. Adicionalmente, ao realizar o encolhimento dos coeficientes, a técnica ataca um de suas principais consequências, que é a estimação de variâncias elevadas.

Antes de discutir o procedimento de encolhimento permitido pelas técnicas, convém discutir formalmente alguns conceitos comumente utilizados nessa literatura:

- **Regularização**: é um procedimento de otimização que impõe uma restrição ao mecanismo dos Mínimos Quadrados Ordinários (*Ordinary Least Squares*. OLS), de maneira a incentivar uma solução caracterizada pela atenuação dos efeitos da complexidade inerente aos modelos de alta dimensão (HD) – por meio ou não de esparsidade. No caso de aplicar soluções esparsas, o faz pela introdução de maior parcimônia na especificação, prevenindo o sobreajuste dos dados. Também conhecido como "método de regressão penalizada", sendo um tipo de Mínimos Quadrados Restritos.

- **Penalização**: é o procedimento utilizado pela regularização, no qual são aplicados valores adicionais às soluções encontradas pelo estimador que envolvem coeficientes de alto valor absoluto, de forma a afastá-las da situação de mínimo desejada. A penalização dessas soluções é realizada por meio de um parâmetro de penalização tipicamente denotado com a letra grega lambda ($\lambda$, parâmetro de ajuste, hiperparâmetro), que controla a intensidade do procedimento. A definição do parâmetro de ajuste é arbitrada *ex-ante*, mas pode ser escolhido um parâmetro ótimo com base em um conjunto de rodadas em uma grade de pesquisa (*grid search*), selecionando-se o parâmetro a partir de critérios de maior precisão de previsões em validações cruzadas, dentre outros. Importante mencionar que, no âmbito da regularização, sendo aplicada uma penalização sobre o coeficiente das variáveis, essas são automaticamente descritas como "variáveis penalizadas" – mais apropriadamente, "variáveis cujo coeficiente está sujeito ao encolhimento". Em alguns arcabouços empíricos, o pesquisador pode selecionar apenas um subconjunto das variáveis para sofrerem esse tipo de penalização, criando-se os subconjuntos das "variáveis penalizadas" e "variáveis não penalizadas". No âmbito do LASSO, algumas das variáveis penalizadas podem ser inativadas, ou seja, ter seu coeficiente reduzido ao valor zero, o que não significa que apenas essas foram "penalizadas". Adicionalmente, no conceito do PDS-LASSO, como veremos mais adiante, algumas das variáveis – penalizadas ou não penalizadas – podem ter seu resultado de estimação pós-dupla seleção sendo não estatisticamente significantes, o que não implica em dizer que foram penalizadas.

---
[2] Os autores se referem a s como uma "esparsidade exata", dado que os coeficientes sempre ou são não nulos, ou são nulos, sem a possibilidade de coeficientes não nulos com magnitude suficientemente "pequeno" – hipótese que também relaxam no trabalho.

- **Pré-padronização**: procedimento prévio de transformação das variáveis do modelo de regressão de maneira que cada variável tenha média igual a zero e desvio padrão igual a 1. Procedimento imprescindível para a regressão penalizada, dado que a restrição inserida na extração do ótimo envolve lidar com os valores dos coeficientes, que flutuam ao sabor da escala de cada variável. Com a pré-padronização, garante-se que a escala foi removida e os coeficientes são comparáveis e tratados de maneira equânime. Como dizem Ahrens, Hansen & Schaffer (2020, p. 180) "*Ao contrário do OLS, o LASSO não é invariante às transformações lineares, por isso* [remover] *a escala é importante*". Sem a pré-padronização, a escala dos coeficientes iria interferir espuriamente na solução obtida pela regularização. Nas palavras de Tibishirani (1997, p. 394, tradução livre): "*O método lasso requer padronização inicial dos regressores, de modo que o esquema de penalização seja justo para todos os regressores.*" Inclusive, o autor naquele estudo recomenda padronizar também as variáveis dummy, muito embora reconheça que a escala relativa entre variáveis contínuas e categóricas neste esquema pode ser um tanto arbitrária, perdendo-se a interpretação direta dos coeficientes.

- **Encolhimento**: redução, pela regressão regularizada, dos valores absolutos dos coeficientes estimados das variáveis pré-padronizadas, fruto do procedimento de penalização.

- **Seleção de modelo**: resultado adicional da regressão penalizada realizada pelo LASSO, onde apenas um subconjunto dos coeficientes originais mantém-se não nulo – apesar de com valores absolutos inferiores –, enquanto os demais tornam-se nulos. A "seleção", nesse caso, se dá por inativação (anulação) de um conjunto de coeficientes, de forma que as respectivas variáveis são consideradas fora do modelo final, selecionado.

- **Inativação**: também conhecida como "exclusão do conjunto ativo", ou "dispensa da variável" (Tibshirani, 2013, p. 1481); consistente com a terminologia de "conjunto ativo" de Ahrens, Hansen & Schaffer (2020), a inativação significa que a variável, embora inicialmente pertencente ao conjunto de regressores sob consideração, não mais pertence ao modelo final, por ter tido o seu coeficiente zerado no processo de penalização realizado pelo LASSO. No procedimento de regularização, todos os coeficientes são penalizados de forma a encolher. Entretanto, alguns deles permanecem no conjunto ativo ("variáveis ativas"), enquanto que uma grande parte deles é inativada ("variáveis inativas"), produzindo-se uma solução esparsa – isto é, com vários coeficientes zerados. Penalização não implica necessariamente em inativação.

Observam Ahrens, Hansen & Schaffer (2020), que o objetivo principal da regressão regularizada, assim como os métodos de aprendizado de máquina supervisionados de maneira mais geral, é a previsão, e não interpretação. "*A regressão regularizada normalmente não produz estimativas que podem ser interpretadas como causais, e a inferência estatística sobre esses coeficientes é complicada. Embora a regressão regularizada possa selecionar o modelo verdadeiro à medida que o tamanho da amostra aumenta, geralmente esse é o caso apenas sob suposições sólidas. No entanto, a regressão regularizada pode auxiliar na inferência causal sem depender das fortes suposições necessárias para a seleção perfeita do modelo*" (Ahrens, Hansen & Schaffer, 2020, p. 177)

Usaremos, na presente exposição, as apresentações de Ahrens, Hansen & Schaffer (2020), Tibshirani (1996), Hastie, Tibshirani & Wainwright (2015), e Bühlmann & de Geer (2011). Considere um modelo linear de alta dimensão, HD:

$$y_i = \beta_0 + \sum_{j=1}^{p} \beta_j x_{ji} + \varepsilon_i \qquad (8)$$

onde, $y_i$ é o regressando, referente à observação $i$, $i = 1, ..., n$; $x_{ji}$ é o $j$-ésimo preditor; o pesquisador se defronta com a tarefa de estimar $p$ parâmetros ($\beta_1, \beta_2, ..., \beta_p$) a partir de uma base de dados; o número de variáveis preditoras, $p$, é grande. A situação mais comum em regressão linear é que as variáveis possuam escala própria e não tenham variância igual. Nesse caso, a abordagem mais comumente utilizada é pré-padronizar os dados, de maneira que:

$$\text{Média}(x_j) = 0, \ Var(x_j) = 1, \ \forall j. \tag{9}$$

A pré-padronização do regressando (y) não é necessária, mas pode ser centralizada – ou seja, removida a média, de forma que a nova média seja igual a zero – para retirar o intercepto da análise. Dessa forma, podemos considerar:

$$y_i = \sum_{j=1}^{p} \beta_j x_{ji} + \varepsilon_i \tag{10}$$

O problema clássico de estimação pelo método dos Mínimos Quadrados Ordinários (Ordinary Least Squares, OLS) pode ser descrito da seguinte forma. A solução para estimação dos parâmetros oferecida pelo OLS é a seguinte:

$$\hat{\beta}_{OLS} = \arg\min_{b} \sum_{i=1}^{N} \left( y_i - \sum_{j=1}^{p} b_j x_{j,i} \right)^2, \tag{11}$$

onde $\sum_{i=1}^{N} \left( y_i - \sum_{j=1}^{p} b_j x_{j,i} \right)^2$ é a Soma dos Quadrados dos Resíduos (RSS, *Residual Sum of Squares*), $\beta = \langle \beta_1 \ \beta_2 \ ... \ \beta_p \rangle$ é o vetor de parâmetros desconhecidos, $\hat{\beta}_{OLS}$ é o vetor de coeficientes estimados pelo OLS na solução ótima, e $b$ rastreia a gama de vetores de coeficientes possíveis.

A solução oferecida pelas regressões LASSO e RIDGE envolvem adicionar um termo de penalização à solução dos mínimos quadrados que implique em coeficientes estimados elevados. Conforme discorrem Ahrens, Hansen & Schaffer (2020), uma "regressão linear regularizada" é uma extensão direta da regressão linear. Da mesma forma como o OLS, a regressão linear regularizada minimiza a soma dos desvios quadrados entre os valores observados e preditos pelo modelo, mas impõe uma penalidade de regularização destinada a limitar a complexidade do modelo, introduzindo um pouco de viés para reduzir sua variância. No caso do LASSO, a penalização se dá na forma de uma restrição sobre a somatória dos valores absolutos dos coeficientes das variáveis, a chamada "Regularização $\ell_1$" (ou "Penalização $\ell_1$"):

$$\hat{\beta}_{LASSO}(\lambda) = \arg\min_{b} \sum_{i=1}^{N} \left( y_i - \sum_{j=1}^{p} b_j x_{j,i} \right)^2 + \lambda \sum_{j=1}^{p} |b_j| \tag{12}$$

onde o $\hat{\beta}_{LASSO}$ é o vetor de coeficientes estimados pelo LASSO na solução ótima, o termo $\sum_{j=1}^{p} |b_j|$ diz respeito à soma dos valores absolutos (módulo) dos coeficientes e $\lambda > 0$ é um parâmetro de ajuste, de intensidade da penalização, e que irá promover o encolhimento dos coeficientes.

No caso da regressão RIDGE, temos uma restrição sobre a somatória dos quadrados dos coeficientes das variáveis, a chamada "Regularização $\ell_2$" (ou "Penalização $\ell_2$"):

$$\hat{\beta}_{RIDGE}(\lambda) = \arg\min_{b} \sum_{i=1}^{N} \left( y_i - \sum_{j=1}^{p} b_j x_{j,i} \right)^2 + \lambda \sum_{j=1}^{p} b_j^2 \tag{13}$$

onde o $\hat{\beta}_{RIDGE}$ é o vetor de coeficientes estimados pelo RIDGE na solução ótima, e o termo $\sum_{j=1}^{p} b_j^2$, a soma dos valores ao quadrado dos coeficientes é valor que será encolhido pela penalização com uso de $\lambda$. Note que, no caso do RIDGE, a penalização é ainda mais punitiva de valores altos anormais de $b_j$, elevando qualquer aumento de coeficiente ao quadrado.

Ahrens, Hansen & Schaffer (2020, p. 182) apresentam os clássicos diagramas das soluções geométricas obtidas por meio do LASSO e do RIDGE, os quais reproduzimos em uma versão anotada na Figura 2. Nela, é possível observar os distintos comportamentos de obtenção de solução restrita proporcionados pelas regressões RIDGE e LASSO. As curvas de nível em torno da solução OLS mostram as possíveis soluções fora da situação irrestrita de mínimos quadrados. As soluções restritas serão alcançadas em pontos de toque entre as curvas de nível e as respectivas área de restrição – círculo no caso de RIDGE e losango no caso de LASSO. Apesar de ambas as soluções permitirem o encolhimento dos coeficientes na solução restrita, apenas o LASSO proporciona uma solução onde um conjunto de coeficientes pode ser zerado de fato – no caso da figura, $\beta_2$. Assim, o LASSO permite obtermos uma "solução esparsa", ou seja, uma solução onde coeficientes são nulos no modelo final. O encolhimento permitido por ambas as técnicas, entretanto, possibilita a redução da variância associada aos modelos complexos, dotados de muitas dimensões, dado que o encolhimento reduz a sensibilidade do modelo aos parâmetros.

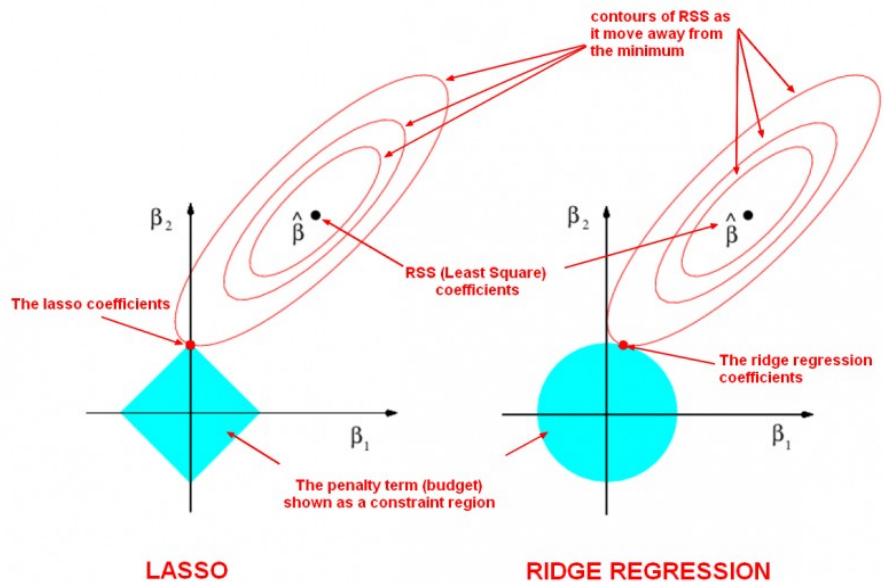

**Figura 2 – Comportamento das penalizações $L_1$ e $L_2$**
(extraído de "Machine Learning", ethen8181.github.io/machine-learning/regularization/regularization)

Uma forma de visualizar o comportamento dos parâmetros estimados na medida em que se aplica uma penalização de intensidade ($\lambda$) crescente é o diagrama de "Trajetória de Regularização" (*Regularization Path*). Na Figura 3 é possível ver como os parâmetros estimados pelo LASSO são zerados na medida em que a penalidade $\ell_1$ se intensifica, o que não acontece com a penalidade $\ell_2$, que os encolhe sem zerar.

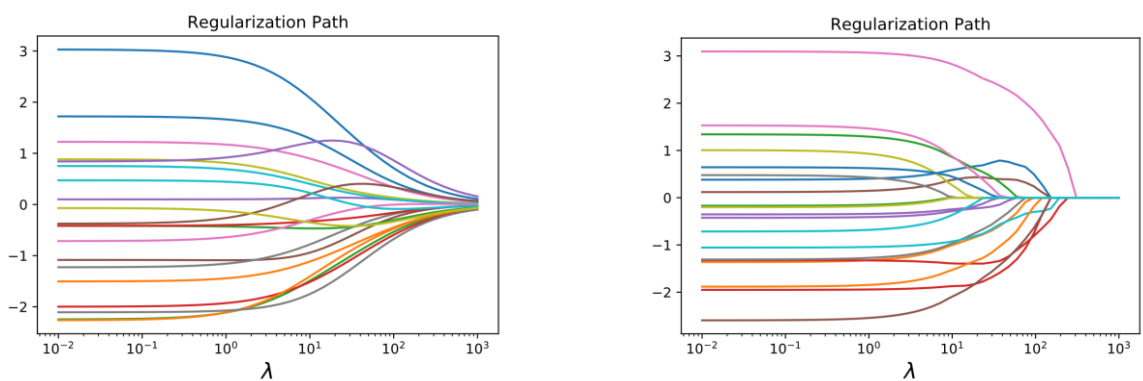

**Figura 3 – Trajetórias de Regularização (Regularization Paths) – RIDGE e LASSO**
(extraído de cvxpy.org/examples/machine_learning/ridge_regression
e cvxpy.org/examples/machine_learning/lasso_regression)

Como discutem Ahrens, Hansen & Schaffer (2020), é notório que os métodos de regressão penalizada dependem fortemente dos parâmetros de ajuste para controlar o grau e o tipo de penalização. No caso do parâmetro $\lambda$, temos que, ao defini-lo igual a zero corresponde a nenhuma penalização, sendo equivalente ao resultado do OLS, enquanto um valor suficientemente grande de $\lambda$ produzirá um modelo vazio, onde todos os coeficientes são nulos, conforme visto nas trajetórias de regularização.

Tibshirani (1996) observa duas vantagens principais do LASSO em relação ao OLS. Primeiro, o LASSO serve como uma técnica de seleção de modelo – coeficientes zerados – e, portanto, facilita a interpretação dos modelos. Em segundo lugar, o LASSO pode superar o OLS em precisão de predição por causa do *tradeoff* viés-variância. Com relação à regressão RIDGE, ao contrário do LASSO, esta não realiza seleção de variáveis – também conhecido como "*model selection*", "*feature selection*" e, em alguns casos da literatura, inserido em contextos como uma ferramenta na área de Extração do Conhecimento, "*Knowledge Discovery in Databases*", KDD, como um meio de se ganhar *insights* sobre os fenômenos (Pantazis et al, 2018). Ahrens, Hansen & Schaffer (2020), observam que a regressão RIDGE não se baseia no pressuposto de esparsidade, o que a torna atrativa nos casos em que esse pressuposto não parece plausível, situação denominada de presença de sinais densos nos dados. Observam os autores que "*problemas densos de alta dimensão são mais desafiadores do que problemas esparsos*" (p. 180). Com relação às propriedades das soluções, a solução proporcionada pela regressão RIDGE tem a garantia de unicidade – como uma norma $\ell_2$ –, enquanto a da regressão LASSO – norma $\ell_1$ –, não possui essa garantia – Pantazis et al (2018). Tibshirani (2013) demonstra que o LASSO possui a propriedade de existência – há ao menos uma solução, dado se tratar de um problema de minimização convexo (p. 1460). O autor também discute a existência de sua solução, apresentando um resultado existente na literatura de que, apesar de não ser uma garantia universal, ela é possível sempre que as colunas de regressores *X* estiverem no que ele denomina de "posição geral" (*general position*). Na prática, sempre que a matriz de preditores forem conjuntamente distribuídas como uma distribuição de probabilidades contínuas, tem-se condições suficientes para a unicidade (pp. 1463 e 1481), condição que se aplica a outros problemas de regularização.

Uma outra propriedade importante da solução LASSO é a seguinte. De acordo com Tibshirani (2013), a partir do desenvolvimento de um método simples para calcular limites marginais inferior e superior para os coeficientes de suas soluções, é possível mostrar que, mesmo em situações de multiplicidade, há a garantia de que os sinais dos coeficientes não serão trocados em soluções alternativas. Assim, demonstra o autor que "*duas soluções LASSO não podem exibir sinais diferentes para uma variável ativa comum, o que implica que os intervalos delimitadores* [dos parâmetros] *não podem conter zero em seus interiores*" (p. 1481, tradução livre).

## VII. CLASSIFICAÇÃO DE REGRESSORES PARA FINS DE ESPECIFICAÇÃO

Uma discussão inicial e muito relevante em econometria diz respeito ao conceito platônico de "Processo Gerador dos Dados, PGD" (*Data-Generating Process*), também chamado de Processo Gerador do Mundo Real" (*Real-World Generating Process*) de uma variável a ser explicada. Davidson & MacKinnon (2003) discorrem que em estatística é comum falarmos da "população" da qual uma amostra é retirada. Explicam os autores, que a expressão "média populacional" é um resquício da época em que a estatística era praticamente apenas a bioestatística, e o objeto de investigação era focado na população humana, em geral de uma cidade ou país específico, da qual amostras aleatórias eram retiradas por estatísticos para estudo. O "peso médio" de todos os membros da população, por exemplo, poderia ser aproximado a partir de uma estimativa da média dos pesos dos indivíduos da amostra – isto é, pela média da amostra dos pesos dos indivíduos. A média da amostra seria, nesse caso, uma estimativa da média da população – considerando a premissa de que a amostra é representativa da população da qual foi retirada. Dizem Davidson & MacKinnon (2003, p. 32): "*Em econometria, o uso do termo população é simplesmente uma metáfora. Um conceito melhor é o de um processo de geração de dados. Por esse termo, queremos dizer qualquer mecanismo que esteja em funcionamento no mundo real da atividade econômica dando origem aos números em nossas amostras, isto é, precisamente o mecanismo que nosso modelo econométrico supostamente descreve*". E complementam: "*O PGD é, portanto, o análogo em econometria de uma população em bioestatística. As amostras podem ser retiradas de um PGD da mesma forma que podem ser retiradas de uma população. Em ambos os casos, as amostras são consideradas representativas do PGD ou da população da qual são retiradas*" (p. 32).

Sendo o o "verdadeiro" mecanismo que gerou os dados de uma base, há que se considerar que a maioria dos PGDs para fenômenos a serem estudados é provavelmente muito complexa, sendo que "*os economistas portanto não pretendem entender todos os detalhes deles*" (Davidson & MacKinnon, 2003, p. 32). Pelo contrário, em teoria da econometria, uma área de investigação fundamental é acerca das propriedades estatísticas dos estimadores, com uso de técnicas de simulação Monte Carlo, onde quase sempre necessário assumir que o PGD é bastante simples. Entretanto, conforme discutimos, nos tempos atuais, onde proliferam os dados em alta dimensão, torna-se cada vez mais difícil trabalhar em modelos econométricos assumido hipóteses demasiadamente simplificadoras com relação ao PGD. Os pesquisadores buscam cada vez mais obter formas de se incorporar a complexidade não observada dos fenômenos a seus modelos, ou ao menos controlar seus efeitos, mas sem incentivar o sobreajuste dos dados.

Com a ideia de um PGD em situações de dados HD, torna-se interessante deixar claro os tipos de variáveis que estamos trabalhando. Essa tarefa de "rotulação de regressores" em diferentes tipos não é imprescindível, mas colabora na melhor compreensão das possibilidades dos estimadores e rotinas computacionais disponíveis para a redução da dimensionalidade dos modelos. Utilizaremos alguns desses rótulos, listados a seguir, incorrendo no risco de abusar da ideia.

## VII.1. Quanto à participação efetiva no PGD

Essa distinção utiliza categorização aplicada na literatura dos estimadores por média de modelos ("*Model Averaging*"), advinda de Danilov & Magnus (2004):

- variáveis "focais" (*focus variables*), que também podem ser chamadas de variáveis "de geração": são variáveis que são sempre incluídas no modelo, pois devem provavelmente pertencer ao PGD por alguma razão teórica ou de outra esfera, como estudos anteriores e evidência anedótica – digamos que a grande maioria dos especialistas no fenômeno não teria dúvida quanto a esse pertencimento; Danilov & Magnus (2004) explicitamente dizem que são "variáveis explicativas que queremos no modelo (...) independentemente dos valores $t$ encontrados dos parâmetros $\beta$ (...)" (p. 29). Variáveis focais em princípio não devem penalizadas em procedimentos de regularização de parâmetros.

- variáveis "auxiliares" (*auxiliary variables*), ou variáveis "de experimentação", "de teste": são variáveis explicativas adicionais às variáveis focais, e que se tem dúvida quanto à sua inclusão no modelo, ou seja, quanto ao seu pertencimento no PGD. Em geral interessa a inferência quanto a essa variável por se tratar de uma variável de interesse para o estudo, e um teste de hipótese da sua significância estatística é o que se procura primariamente realizar. Mas também podem ser variáveis utilizadas em experimentos de verificação de robustez de um modelo principal onde não estão contidas. Por princípio de conservadorismo da análise, variáveis de experimentação devem sempre ser expostas à penalização/seleção de modelos, no caso de procedimentos de regularização de parâmetros.

Uma característica tanto de variáveis focais quanto auxiliares é que ambas tipicamente têm os seus coeficientes estimados reportados nas tabelas de resultados dos estudos. Em contraste, como veremos abaixo, existem variáveis "de incômodo" que tipicamente não temos interesse em interpretar seus coeficientes estimados, e, portanto, são omitidas das tabelas de resultados. Adicionalmente, ao conjunto união de variáveis focais e variáveis auxiliares denominamos "variáveis de baixa dimensão", como também discutiremos na sequência.

## VII.2. Quanto à função e interpretação no modelo

- variável "de interesse", "de políticas", "de tratamento": esses regressores em geral são variáveis auxiliares (para testar se uma política ou tratamento de fato produziu efeitos), mas também podem ser variáveis focais – quando se tem certeza dos seus impactos, mas busca-se quantificar a intensidade de seus efeitos, ou mesmo analisar seu sinal. Importante enfatizar que, em um estudo de inferência de causalidade – isto é um estudo econométrico de interpretação de relações, e não de previsão –, a(s) variável (variáveis) de interesse são em geral regressoras, não se confundindo com os regressandos, que são as variáveis nas quais desejamos decompor os efeitos por meio da regressão.

- variáveis "terceiras" ou "fatores de confusão" (*confounding factors*): são fatores não observados, latentes, e subjacentes ao problema investigado – ou seja, integram o PGD –, e que em que podem ser, ou em geral são, correlacionados com a(s) variável(is) independente(s) de interesse. Podem ser, por exemplo, características invariantes no tempo não-mensuradas dos indivíduos de um painel ou cross-section de participantes em um estudo, ou qualquer influência não intencional do pesquisador, em uma manipulação experimental (Salkind, 2010), ou dos indivíduos sob análise, ou mesmo de fatores outros não antecipados. Haig (1992) descreve que essas são "variáveis terceiras", pois interferem na relação de causalidade entre X e Y, mas que, por serem correlacionadas com X e Y, mas não observáveis, trazem ruído na estimação do efeito de X em Y.

- controles ou "variáveis de incômodo" (*nuisance variables*): são variáveis *dummy*, *proxies* ou outros artifícios que o pesquisador insere na equação para endereçar a questão da existência de variáveis terceiras no PGD. São denominadas de "*nuisance*" ("de incômodo") na medida em que devem ser controladas estatisticamente de alguma forma no estudo, tão somente para viabilizar a inferência de causalidade das variáveis de interesse, sendo os seus parâmetros estimados ("*nuisance parameters*", como em Basu, 2011) em geral de interpretação com pouco interesse direto ou mesmo prejudicada. Por isso tipicamente os coeficientes estimados dessas variáveis não são reportados nas tabelas de resultados dos estudos – ou seja, são omitidos nos artigos. Em geral, qualquer parâmetro que interfere na análise de outro, mas que a informação obtida diretamente dele é sem relevância direta para o problema, pode ser considerado um "parâmetro de incômodo" (Basu, 2011, p. 279). Uma variável também pode deixar de ser "*nuisance*" caso se torne objeto de estudo. Quando o controle estatístico de variáveis terceiras não é realizado, ou é efetuado de maneira insatisfatório, em geral encontra-se na literatura a denominação de "*confounding variable*" (variável de confusão de efeitos) para essas variáveis – variáveis latentes que, ao falhar em controlar, prejudicam a validade interna de um experimento e/ou as análises de relação de causalidade e impossibilitando inferências *ceteris paribus* (efeitos isolados de uma variável).

## VII.3. QUANTO À DIMENSIONALIDADE

- variáveis de alta dimensão (*high-dimension variables*), também chamadas de "controles de alta dimensão": o mesmo que "controles" ou "*nuisance variables*", para o caso de um problema de alta dimensionalidade. De posse de dados HD, as possibilidades de controles de efeitos não observáveis aumenta consideravelmente, por meio de *dummies* de agrupamentos de indivíduos semelhantes, efeitos fixos de indivíduos em uma estrutura de painel de dados, efeitos temporais, ou outras formas de controle que são específicas de cada caso. O número de controles utilizados pode ser elevado, a depender da complexidade do fenômeno – e por isso surge a necessidade de regularização –, mas pelo fato de haver a possibilidade de serem incluídos no estudo e permitirem o controle das variáveis terceiras, melhoram qualitativamente as inferências estatísticas – o que é uma das vertentes da "bênção da dimensionalidade".
- variáveis de baixa dimensão (*low-dimension variables*): no caso de um problema de alta dimensionalidade, toda variável que não se enquadra como variável de incômodo ("*nuisance*") é uma variável de "baixa dimensão", podendo ser focal ou auxiliar. Assim ao conjunto união de variáveis focais e variáveis auxiliares, em geral estamos nos referindo ao conjunto completo de "variáveis de baixa dimensão". Em suma, essas são variáveis de especificação do modelo, e que cuja estimação será beneficiada com a inclusão dos controles HD.

## VII.4. QUANTO À REGULARIZAÇÃO

- variáveis penalizadas: variáveis cujo coeficiente está sujeito ao procedimento de encolhimento pela imposição da restrição de regularização.
- variáveis não penalizadas: em alguns arcabouços empíricos, o pesquisador pode configurar o problema de modo a selecionar apenas um subconjunto das variáveis existentes para sofrerem a penalização da regularização. Cria-se, assim, subconjuntos das "variáveis penalizadas" e "variáveis não penalizadas".

## VII.5. QUANTO À INCLUSÃO NO CONJUNTO ATIVO DO LASSO

- variáveis ativas: variáveis mantidas no "conjunto ativo" de variáveis – ou seja, as variáveis penalizadas cujo coeficiente estimado é não nulo após a regularização. Inclui também as variáveis que foram não penalizadas, seja por decisão do pesquisador, ou por necessidade do modelo – por exemplo, as variáveis endógenas em um procedimento IV-LASSO são sempre ativas, como veremos mais adiante.
- variáveis inativadas: No âmbito do LASSO, algumas das variáveis penalizadas podem ser inativadas, ou seja, ter seu coeficiente reduzido ao valor zero, sendo, assim, equivalente a dizer que foram variáveis "não selecionadas" pelo LASSO. Assim, temos o resultado adicional da regressão penalizada realizada pelo LASSO, onde apenas um subconjunto dos coeficientes originais mantém-se não nulo – apesar de com valores absolutos inferiores –, enquanto os demais tornam-se nulos. A "seleção", nesse caso, se dá por inativação (anulação) de um conjunto de coeficientes, de forma que as respectivas variáveis são consideradas fora do modelo final, selecionado. A inativação é também conhecida como "exclusão do conjunto ativo", ou "dispensa da variável" (Tibshirani, 2013, p. 1481); consistente com a terminologia de "conjunto ativo" de Ahrens, Hansen & Schaffer (2020), a inativação significa que a variável, embora inicialmente pertencente ao conjunto de regressores sob consideração, não mais pertence ao modelo final, por ter tido o seu coeficiente zerado no processo de penalização realizado pelo LASSO. No procedimento de regularização, todos os coeficientes são penalizados de forma a encolher. Entretanto, alguns deles permanecem no conjunto ativo ("variáveis ativas"), enquanto que uma grande parte deles é inativada ("variáveis inativas"), produzindo-se uma solução esparsa – isto é, com vários coeficientes zerados. Penalização não implica necessariamente em inativação. Terem sido inativadas não significa que apenas essas variáveis foram "penalizadas". Outro ponto a se enfatizar é que, como veremos mais adiante, no conceito do PDS-LASSO algumas das variáveis – penalizadas ou não penalizadas – podem ter seu resultado de estimação pós-dupla seleção sendo não estatisticamente significantes, o que não implica em dizer que foram penalizadas; a rejeição no teste de hipóteses desenvolvido na pós-seleção não significa penalização – procedimento esse feito antes desse passo.

## VIII. Procedimentos de estimação pós-LASSO

Os procedimentos de estimação pós-LASSO aqui apresentados têm por objetivo oferecer métodos robustos para inferência sobre o efeito de uma variável de efeito de tratamento (genericamente, uma variável de interesse do estudo) na presença de muitos regressores em um modelo com resíduos possivelmente não gaussianos e heteroscedásticos. Nessa análise, os autores permitem que o número de regressores seja maior do que o tamanho da amostra – modelo de alta dimensionalidade (HD). Para tornar a inferência informativa viável, assumem que o modelo seja aproximadamente esparso – o princípio da esparsidade; isso, segundo os autores, é equivalente a possibilitar que o efeito dos fatores de confusão (variáveis terceiras, *confounding factors*) possa ser controlado com apenas um pequeno erro de aproximação, incluindo um número relativamente pequeno dessas variáveis cujas identidades são desconhecidas – isto é, não observadas. Explicam os autores que a possibilidade de controlar essas variáveis dessa maneira torna possível a estimação do efeito do tratamento (variável de interesse) selecionando aproximadamente o conjunto correto de regressores. No que se segue, discutiremos esses procedimentos com mais detalhe.

Na área de aplicações em LASSO em problemas econômicos, um conjunto de autores norte-americanos produziu uma sequência de publicações de artigos científicos sobre o tema ao longo dos anos 2010. Em particular, na área de estimação pós-lasso, destacamos os autores:

- Alexandre Belloni (Duke University);
- Christian B. Hansen (University of Chicago); e
- Victor Chernozhukov (MIT).

Os estudos desses autores focam no procedimento de regularização por Lasso, utilizando metodologias que visam manter sua vantagem principal, referente à redução da dimensionalidade com seleção de modelos, mas reduzir o viés inerente a esse procedimento – problema que levanta preocupações quanto à qualidade da inferência estatística de causalidade permitida pelos modelos. Destaca-se, como textos de embasamento teórico na área, os seguintes artigos:

- Pós-Dupla Seleção por Lasso (*Post-Double Selection*, denotado como "PDS-LASSO" nas rotinas do *lassopack,* um pacote de regressão regularizada desenvolvido para o software Stata, usado para esse tipo de estimação).
    - citação: Belloni, Chernozhukov e Hansen (2014a);
    - artigo "*Inference on treatment effects after selection among high-dimensional controls*". Iremos nos referir a esse artigo como "BCH14";
    - publicado na *The Review of Economic Studies*.
- Pós-Regularização (*Post-Regularization*, que indicaremos como "PR-LASSO", denotado como "Metodologia CHS" no *lassopack*).
    - citação: Chernozhukov, Hansen e Spindler (2015);
    - artigo "*Post-selection and post-regularization inference in linear models with many controls and instruments*". Iremos nos referir a esse artigo como "CHS15";

- publicado na *American Economic Review*.
- LASSO com Variáveis Instrumentais (*Instrumental Variables LASSO*, denotado como "IV-LASSO" nas rotinas do *lassopack*).
  - citação: Belloni, Chen, Chernozhukov & Hansen (2012);
  - artigo "*Sparse models and methods for optimal instruments with an application to eminent domain*". Iremos nos referir a esse artigo como "BCCH12";
  - publicado na *Econometrica*.

Antes de prosseguirmos, vale uma nota: no presente trabalho, usamos as siglas BCH14, CHS15 e BCCH12 apenas para facilitar a menção aos estudos da área, dado que os autores foram profícuos em produzir artigos científicos publicados e não-publicados – o que pode ser confuso para acompanhar. Entretanto, essas siglas não são utilizadas na literatura da área, servindo apenas de finalidade didática na exposição dos trabalhos.

## VIII.1. ARCABOUÇO GERAL

Para melhor descrever o problema, utilizaremos a estrutura geral apresentada em BCH14. Começando com a definição do tipo de modelo de regressão. BCH14 assumem um "modelo parcialmente linear" (*partially-linear model*, PLM). Segundo Härdle, Liang & Gao (2012) "*modelos parcialmente lineares*" *(PLM) são "modelos de regressão nos quais a resposta depende de algumas covariáveis linearmente, mas de outras covariáveis não parametricamente.*" (abstract). Dizem os autores que os PLMs são modelos que generalizam as técnicas de regressão linear padrão. Os PLM são mais flexíveis que o modelo tradicional, dado que, sendo semiparamétricos – ou seja, contêm tanto componentes paramétricos quanto não paramétricos –, permitem que o fenômeno a ser modelado contenha relações não lineares com variáveis que não são observáveis, ou afetam o problema de uma forma funcional não observável ou não modeláveis da forma padrão. Assuma, por exemplo, o seguinte modelo parcialmente linear:

$$y_i = \alpha d_i + g_0(x_i) + \varepsilon_i \tag{14}$$

$$d_i = m_0(x_i) + u_i \tag{15}$$

onde $y_i$ é a variável dependente, $i$ representa a $i$-ésima observação, $d_i$ é uma variável de interesse e $x_i = (x_1, \dots, x_p)$ é um vetor $p$-dimensional que engloba todos os controles, podendo ser de alta dimensão. $g_0(x_i)$ é uma função dos controles e $\varepsilon_i$ é o termo de erro aleatório. A grande questão relacionada ao PLM é que $g_0(x_i)$ é uma função desconhecida, ou seja, uma parte do processo gerador dos dados de $y_i$ que o pesquisador tem que aproximar de alguma forma, por exemplo, com o uso de variáveis de controle.

Importante observar que a componente de interesse em (14) não precisa ser necessariamente um escalar, tendo sido definido dessa forma apenas para simplificar a exposição. Ao contrário, pode haver uma especificação mais completa dessas variáveis de "baixa dimensão" – *low dimension variables*, variáveis indispensáveis que estariam em um modelo de baixa dimensão de qualquer forma. Com a presença da componente desconhecida de alta dimensão $g_0(x_i)$, estimar os parâmetros da(s) variável (variáveis) de interesse torna-se o grande desafio. Deseja-se tecer inferências estatísticas sobre o efeito *ceteris paribus* de $d$ em $y$, sendo assim $\alpha$ o coeficiente mais importante que procuramos estimar. A componente não-paramétrica $g_0(x_i)$ modela como "fatores de confusão" (*confounding factors*) surgem no problema. Esses fatores de confusão são em geral fatores latentes, não observáveis, e que são possivelmente correlacionados com $d_i$. A segunda equação modela justamente a dependência da variável de interesse com relação aos controles, como em BCH14 e Chernozhukov et al (2018). Comentam Chernozhukov et al (2018), que "*esta equação não é de interesse per se, mas é importante para caracterizar e remover o viés de regularização*" (p. C2). Os fatores de confusão $x_i$ afetam tanto a variável de interesse $d_i$ por meio da função $m_0(x_i)$ e a variável dependente principal $y_i$ por meio da função $g_0(x_i)$. Definem os autores como espaço dos "parâmetros de incômodo" (*nuisance parameters*) sendo $\eta_0$, igual a:

$$\eta_0 = (m_0, g_0) \tag{16}$$

Ao utilizar controles, a proposta de um modelo em alta dimensão seria aproximar à função $g_0(x_i)$ – e a função $m_0(x_i)$ – por meio de um elevado conjunto de variáveis – nesse caso denominadas de "variáveis de incômodo" (nuisance variable), que possibilitarão a estimativa de "parâmetros de incômodo" (*nuisance parameters*). A esse conjunto de parâmetros, soma-se os parâmetros da função $m_0(x_i)$.

Assim, no exemplo dado por BCH14 podemos perceber que, em uma regressão contendo uma variável interesse (ou mais delas), mas em um arcabouço marcado pela alta dimensionalidade, a existência de um conjunto alto de fatores não-observáveis a serem controlados nessa regressão torna o problema mais desafiador. Sendo eles correlacionados com a variável de interesse, temos um problema a ser endereçado pelo procedimento de estimação. Não controlar esses fatores na regressão principal geraria um grande problema de viés de variáveis omitidas, em particular por endogenia devido à correlação com esses fatores de confusão. A proposta de um modelo econométrico em alta dimensão que desejasse minimizar o problema de variáveis omitidas seria a de controlar esses fatores de confusão diversos tipos de controles. Dentre esses controles, podemos listar:

- Efeitos fixos das unidades de painel, para controlar fatores específicos individuais que sejam invariantes ao tempo – procedimento clássico do estimador de efeitos fixos, em geral não sujeito a penalização por LASSO;
- Efeitos temporais, para controlar fatores específicos de cada período de tempo de um painel ou dados agrupados, para controlar as mudanças coletivas das unidades do painel;
- variáveis *proxies* para efeitos não observáveis de fenômenos mais complexos;
- variáveis defasadas ou em diferenças, utilizando a dimensão temporal e/ou espacial;
- variáveis de *one-hot encoding* (*dummies*) de categorias representativas das unidades de painel e seus possíveis agrupamentos;
- termos de ordem superior e interações de todas as variáveis acima listadas.

A estimação de parâmetros dessas diversas variáveis não seria de interesse direto do pesquisador - por isso são chamados de "*nuisance parameters*" -, mas um recurso para aproximar a uma função desconhecida, não-paramétrica, e assim controlar efeitos não observáveis para viabilizar uma identificação consistente do efeito da variável de interesse. Por serem variáveis de alta dimensão, algum procedimento de redução da dimensionalidade deve ser aplicado aos controles utilizados, sendo que a proposta dos autores é o uso de modelos de Pós-Regularização/Seleção do tipo LASSO para se "*procurar fatores de confusão*" nos dados (BCH14, p. 639).

## VIII.2. Pós-Dupla Seleção por LASSO (PDS-LASSO)

Para a descrição do modelo de BCH14, usaremos tanto a discussão apresentada no próprio artigo, quanto a versão simplificada apresentada no site do *lassopack*[3]. Suponha o seguinte problema de estimação, equivalente a (14) e (15):

$$y_i = \alpha d_i + x_i'\beta + \varepsilon_i \tag{17}$$

$$d_i = x_i'\gamma + u_i \tag{18}$$

onde $y_i$ é a variável dependente, $i$ representa a $i$-ésima observação, $d_i$ é uma variável de interesse e $x_i = (x_1, ..., x_p)$ é um vetor $p$-dimensional que engloba todos os controles de alta dimensão, $\beta$ e $\gamma$ são vetores de parâmetros, e $\varepsilon_i$ é o termo de erro aleatório. Novamente, por simplificação utiliza-se apenas uma variável de baixa dimensão – ou seja, uma variável além dos controles, nesse caso a variável de interesse –, o que não precisa ser o caso em aplicações práticas. Importante salientar que o pesquisador não observa/não conhece o conjunto verdadeiro, esparso, de controles e, por isso, têm que encarar o *tradeoff* viés-variância: usar poucos controles ou os controles errados e incorrer no viés de variáveis omitidas; ou usar muitos controles e incorrer em sobreajuste.

Utilizar uma abordagem de estimação em alta-dimensão é relevante nesse caso, pois os controles $x_i$ permitem estimar adequadamente a complexidade do fenômeno subjacente aos dados. Assumindo esparsidade não observável, o pesquisador nesse caso pretende aproximar-se ao máximo da estimação do modelo verdadeiro, mas sem incorrer em variância desnecessária, produzindo modelos pouco generalizáveis. Com base nesse desafio, a principal ideia relativa à abordagem de BCH14 é utilizar o modelo de regularização/seleção LASSO para selecionar quais controles $x_i$ efetivamente utilizar, e assim promover uma estimação consistente de $\alpha$, que é o parâmetro da variável de interesse. Essa abordagem foi denominada de metodologia de "Pós-Dupla Seleção" por LASSO (PDS-LASSO).

---

[3] Ver statalasso.github.io/docs/pdslasso_models.

O algoritmo do PDS-LASSO utiliza três passos, sendo os dois primeiros de regularização/seleção e o último, uma regressão final. O algoritmo é o seguinte (BCH14, p. 610):

- **Passo I (Primeira Seleção de Modelos)**: estimar uma regressão LASSO contendo $y_i$ como a variável dependente e as variáveis de controle $x_i$ como regressores penalizados;
- **Passo II (Segunda Seleção de Modelos)**: estimar uma regressão LASSO contendo $d_i$ como a variável dependente e, novamente, as variáveis de controle $x_i$ como regressores penalizados;
- **Passo III (Estimação Final)**: rodar uma regressão linear contendo $y_i$ como a variável dependente, e, como regressores, $d_i$ e também a união do conjunto de variáveis $x_i$ selecionados nos Passos I e II.

Note que a abordagem PDS-LASSO envolve selecionar, primeiramente, os controles que os dados mostram serem mais relacionados à variável principal $y$. Ou seja, temos o objetivo de redução do problema (e maldição) da dimensionalidade alta na regressão principal nesse primeiro passo – o que seria de se esperar de um procedimento de uma etapa utilizando LASSO, para encontrar o balanceamento ótimo do binômio viés-variância. Entretanto, a metodologia introduz um segundo passo, onde se busca encontrar os controles relevantes explicativos – ou simplesmente correlacionados – com a variável de interesse, $d$, também utilizando regularização. Esse segundo passo possui a interpretação de que a primeira regularização não seria suficiente para encontrar os fatores de confusão que permitiriam estimar de maneira consistente o efeito *ceteris paribus* da variável de interesse na equação principal (17). Assumindo em (18) que $d$ sofre de multicolinearidade com os controles de alta dimensão, pode ocorrer de que o primeiro procedimento não selecione alguns desses controles colineares com $d$. Na omissão desses controles colineares – os não selecionados pelo primeiro LASSO –, não teríamos garantida a inexistência de viés de variável omitida ao rodar a equação principal. A estimação, nesse caso, sofreria de problema de endogeneidade da variável $d$, causando problemas de inferência quanto à causalidade de sua relação com $y$. Sendo assim, o Passo II se faz necessário, onde são identificados, além de um conjunto de controles possivelmente já selecionado no Passo I, mas também um conjunto adicional que havia sido inativado, ou seja, regularizado até ficar nulo na primeira seleção de modelos por LASSO. Por fim, a regressão do Passo III possibilita a geração dos usuais resultados de Mínimos Quadrados – com ou sem outros procedimentos estatísticos possíveis, como controle de heteroscedasticidade, efeitos fixos, uso de variáveis instrumentais, etc. Esses resultados são produzidos com uso apenas do conjunto de controles selecionados nos dois passos anteriores – redução da dimensionalidade. Importante notar que o procedimento final (Passo III) permite a suavização do "viés de atenuação" incorrido pela inserção de viés que a regularização provoca, dado que retomamos uma estimação não penalizada em última instância. Adicionalmente, o procedimento tradicional produz a geração de erros padrões das estimativas e possibilita testes de hipóteses – o que não é possibilitado facilmente pela regressão regularizada.[4]

Em termos mais formais, ao discutir as características de sua metodologia, BCH14 apresentam resultados teóricos a respeito das propriedades do estimador da variável de interesse resultante ($d_i$). Pode-se listar essas características do PDS-LASSO da seguinte forma:

- possibilita a <u>seleção de variável imperfeita em qualquer uma das duas etapas de seleção</u>; enfatizam que a principal característica atrativa do método proposto é que ele permite a seleção imperfeita dos controles;
- permite que os erros sejam não-gaussianos e heteroscedásticos;
- o estimador é "consistente raiz-$n$" (*root-n consistent, ou $\sqrt{n}$-consistent*), o que significa que, mesmo com a incerteza (e possível incorreta) parametrização da componente não-paramétrica do modelo parcialmente linear, a probabilidade de desvio do parâmetro estimado e o parâmetro verdadeiro (viés) obedece uma distribuição de probabilidades de ordem $N^{-1/2}$, ou seja é função do inverso da raiz quadrada de N (tamanho da amostra), tendendo a zero na maioria das aproximações assintóticas (Robinson, 1988).

---

[4] Hastie, Tibshirani & Friedman (2009), p. 12: "*Os erros padrão para as estimativas de mínimos quadrados vêm das fórmulas usuais. Não existe uma fórmula tão simples para o LASSO (..)*". Os autores utilizam *bootstrap* para obter estimativas dos erros padrões. Goeman, Meijer & Chaturvedi (2018, p.18) discutem "*É uma pergunta muito natural pedir erros padrão dos coeficientes de regressão ou outras quantidades estimadas. Em princípio, esses erros padrões podem ser facilmente calculados, e. usando o bootstrap.*" e também "*Quaisquer cálculos baseados em bootstrap podem apenas fornecer uma estimativa da variância das estimativas. Estimativas confiáveis de viés somente estão disponíveis se estimativas confiáveis não enviesadas estiverem disponíveis, o que normalmente não é o caso em situações em que estimativas penalizadas são utilizadas.*"

- possibilita uma <u>inferência estatística</u> que é <u>válida uniformemente</u> – ou seja, fornece intervalos de confiança estáveis – em um número alto de classes de modelos; que "*são válidos uniformemente em uma grande classe de modelos*" (p .608). Em contraste, os estimadores de seleção pós-modelo padrão (de seleção única) falham em fornecer essa "inferência uniforme" e, dessa forma, os testes de hipótese não mantêm uma mesma distribuição de probabilidades assintótica ao longo do espaço de parâmetros, como por exemplo, uma mesma probabilidade assintótica de rejeição da hipótese nula, mesmo em casos simples com um número pequeno e fixo de controles. Aliás, argumentam que esse é o principal resultado teórico do artigo (p. 615, equação 2.10), onde demonstram que o estimador PDS obedece a uma distribuição assintótica que é Normal Padrão, ou seja $\sim N(0,1)$, sob condições de esparsidade aproximadas, uniformemente dentro de um rico conjunto de processos geradores de dados estudados.

- ainda acerca da inferência uniforme, ao realizar a seleção em dois passos – a "dupla seleção" – contribui-se com a redução do viés de variáveis omitidas, sendo assim possível realizar uma inferência de maior qualidade após a seleção do modelo.

- atinge os limites de eficiência semiparamétrica em algumas condições. Newey (1990) define "limites de eficiência", como a perda de eficiência que pode resultar de uma abordagem semiparamétrica, em vez de uma abordagem paramétrica, a ser utilizada em um estudo empírico.

### VIII.2.1. Exemplo empírico de PDS-LASSO (BCH14)

BCH14 utilizam o estudo de Donohue III & Levitt (2001), sobre o efeito do aborto na criminalidade nos Estados Unidos, para motivar a metodologia PDS-LASSO. Reproduzimos abaixo trechos inteiros da discussão dessa ilustração (BCH14, p. 636-640), com a finalidade didática de exposição da abordagem.

> *"O problema básico em estimar o impacto causal do aborto sobre o crime é que as taxas de aborto em nível estadual não são distribuídas aleatoriamente e parece provável que haverá (outros) fatores associados tanto às taxas de aborto quanto de criminalidade. É claro que qualquer associação entre a taxa de aborto atual e a taxa de crime atual provavelmente será espúria. No entanto, mesmo se olharmos para, digamos, a relação entre a taxa de aborto de 18 anos no passado e a taxa de criminalidade entre as pessoas atualmente com 18 anos, a falta de atribuição aleatória torna difícil estabelecer uma relação causal sem controles adequados. Um fator de confusão óbvio é a existência de diferenças persistentes de estado para estado em políticas, atitudes e dados demográficos que provavelmente estão relacionados ao aborto em nível estadual geral e às taxas de criminalidade. Também é importante controlar com flexibilidade as tendências agregadas. Por exemplo, pode ser o caso de que as taxas de criminalidade nacional tenham caído durante algum período, enquanto as taxas de aborto nacionais estavam aumentando, mas essas tendências foram impulsionadas por fatores completamente diferentes. Sem controlar essas tendências, seria errôneo associar a redução da criminalidade ao aumento do aborto. Além dessas diferenças gerais entre os estados federativos e períodos, há outras características que variam ao longo tempo, como renda em nível estadual, policiamento ou uso de drogas, para citar algumas que podem estar associadas ao crime atual e ao aborto anterior."* (BCH14, pp. 636-637, com grifos próprios e tradução livre).

BCH14, em seu exemplo, utilizam o modelo de Donohue III e Levitt (2001), que estimam os determinantes das taxas de criminalidade em nível estadual nos Estados Unidos entre 1985 a 1997. Dizem os autores:

> *"Para lidar com esses fatores de confusão, Donohue III e Levitt (2001) estimam um modelo para as taxas de crimes em nível estadual de 1985 a 1997, no qual eles condicionam vários desses fatores. Sua especificação básica é*
>
> $$y_{cit} = \alpha_c \alpha_{cit} + w_{it}' \beta_c + \delta_{ci} + y_{ct} + \varepsilon_{cit} \qquad \text{(Eq. 6.45)}$$
>
> *onde i indexa estados, t indexa tempos, c ∈ {violento, propriedade, assassinato} indexa o tipo de crime, $\delta_{ci}$ são efeitos específicos estaduais e que controlam quaisquer características específicas do estado invariáveis no tempo, $y_{ct}$ são efeitos específicos que controlam com* *flexibilidade quaisquer tendências agregadas, $w_{it}$ são um conjunto de variáveis de controle para controlar fatores de confusão em nível estadual, $\alpha_{cit}$ é uma medida da taxa de aborto relevante para o tipo de crime c, e $y_{cit}$ é a taxa de criminalidade para o tipo de crime c. Para o conjunto de controles específicos estaduais e que variam ao longo tempo (componente $w_{it}$), Donohue III e Levitt (2001) usam o log de prisioneiros defasados per capita, o log de policiais per capita defasados, a taxa de desemprego, a renda per capita, a taxa de pobreza, a generosidade do AFDC [Aid to Families with Dependent Children, um programa federal] no tempo t-15, uma dummy para a lei de armas ocultas, e o consumo de cerveja per capita.(...)."* (BCH14, p. 637, com grifos próprios e tradução livre e com remoção das notas de rodapé).

BCH14 conferem elevada importância para o procedimento de efeitos fixos de indivíduo e temporais. Adicionalmente, buscam controlar fatores não observáveis por meio da extração da primeira diferença das variáveis:

*"(...) Dada a importância aparentemente óbvia de controlar os efeitos de estado e tempo, consideramos esses efeitos em todos os modelos que estimamos. Optamos por eliminar os efeitos de estado por meio da diferenciação em vez de incluir um conjunto completo de dummies de estado, mas incluir um conjunto completo de dummies de tempo em cada modelo. Assim, estimaremos modelos da forma*

$$y_{cit} - y_{cit-1} = \alpha_c(\alpha_{cit} - \alpha_{cit-1}) + z'_{cit}\kappa_c + g_{ct} + \eta_{cit} \qquad (Eq.\ 6.46)$$

*onde $g_{ct}$ são efeitos de tempo. Usamos os mesmos dados em nível de estado que Donohue III e Levitt (2001), mas excluímos Alasca, Havaí e Washington, D.C., o que dá uma amostra com 48 observações transversais e 12 observações de séries temporais para um total de 576 observações. Com essas exclusões, nossas estimativas de linha de base usando os mesmos controles de (6.45) são bastante semelhantes às relatadas em Donohue III e Levitt (2001). (...)"* (BCH14, p. 637, com grifos próprios e tradução livre e com remoção das notas de rodapé).

Note que os autores mencionam que possuem uma motivação teórico-empírica para usar a diferenciação (primeira-diferença), em vez de níveis. No apêndice eles discutem os resultados utilizando níveis e empregando efeitos fixos.

A discussão mais relevante no que tange o ponto da alta dimensionalidade em BCH14 está na seguinte passagem:

*"Nosso principal ponto de partida de Donohue III e Levitt (2001) é que permitimos um conjunto $z_{cit}$ muito mais rico do que o permitido por $w_{it}$ no modelo (6.45). Nosso $z_{cit}$ <u>inclui termos de ordem superior e interações das variáveis de controle</u> definidas acima. Além disso, colocamos as condições iniciais e as diferenças iniciais de $w_{it}$ e $\alpha_{cit}$, e médias internas estaduais [within-state averages] de $w_{it}$ dentro do nosso vetor de controles $z_{cit}$. Esse acréscimo abre a possibilidade de que possa haver alguma característica de um estado que está associada tanto à sua taxa de crescimento do aborto quanto à sua taxa de crescimento do crime. Por exemplo, ter níveis inicialmente elevados de aborto pode estar associado a altas taxas de crescimento no aborto e baixas taxas de crescimento no crime. A falha em controlar esse fator poderia levar a atribuir erroneamente o efeito desse fator inicial, talvez impulsionado por políticas ou dados demográficos em nível estadual, ao efeito do aborto. Finalmente, permitimos <u>tendências mais gerais</u>, <u>permitindo uma tendência quadrática agregada</u> em $z_{cit}$, bem como <u>interações dessa tendência quadrática com variáveis de controle</u>. Isso nos dá um conjunto de 284 variáveis de controle para selecionar, além dos 12 efeitos de tempo que incluímos em cada modelo."* (BCH14, pp. 638, nota de rodapé 24, com grifos próprios e tradução livre).

Em uma nota de rodapé, os autores esclarecem: "*As identidades exatas dos 284 controles potenciais estão disponíveis mediante solicitação. Consiste em termos lineares e quadráticos de cada variável contínua em $w_{it}$, interações de cada variável em $w_{it}$, níveis iniciais e diferenças iniciais de $w_{it}$ e $\alpha_{cit}$, as médias internas (within) estaduais de $w_{it}$ e interações dessas variáveis com uma tendência quadrática.*" (BCH14, pp. 638, nota de rodapé 24, com grifos próprios e tradução livre).

Com relação ao papel dos controles efetuados, os autores observam que:

> *"Observe que a interpretação das estimativas do efeito do aborto do <u>modelo (6.45) como causal depende da crença [do pesquisador] de que não existem termos de ordem superior das variáveis de controle, nem termos de interação e nem variáveis adicionais excluídas</u> que estão associadas tanto às taxas de crime quanto à taxa de aborto. Assim, <u>controlar um grande conjunto de variáveis conforme descrito acima é desejável do ponto de vista de tornar essa crença mais plausível. Ao mesmo tempo, controlar de forma ingênua diminui nossa capacidade de identificar o efeito da variável de interesse e, portanto, tende a fazer estimativas muito menos precisas</u>."*

Interessante o comentário dos autores sobre os resultados obtidos:

> *"O resultado de estimar o efeito do aborto condicional ao <u>conjunto completo de 284 controles potenciais</u> descritos acima é dado na terceira linha da Tabela 2 [do artigo BCH14]. Como esperado, <u>todos os coeficientes são estimados de forma muito imprecisa</u>. Claro, muito poucos pesquisadores considerariam o uso de 284 controles com apenas 576 observações <u>devido exatamente a esse problema</u>."* (BCH14, pp. 638, nota de rodapé 24, com grifos próprios e tradução livre).

Abaixo inserimos a Tabela 2 de BCH14, para dar uma ideia mais clara dos resultados obtidos pelos autores do estudo:

TABLE 2
*Estimated Effects of Abortion on Crime Rates*

|  | Violent crime | | Property crime | | Murder | |
|---|---|---|---|---|---|---|
|  | Effect | Std. Err. | Effect | Std. Err. | Effect | Std. Err. |
| A. Donohue III and Levitt (2001) Table IV | | | | | | |
| Donohue III and Levitt (2001) Table IV | −0.129 | 0.024 | −0.091 | 0.018 | −0.121 | 0.047 |
| First-difference | −0.152 | 0.034 | −0.108 | 0.022 | −0.204 | 0.068 |
| All controls | 0.014 | 0.719 | −0.195 | 0.225 | 2.343 | 2.798 |
| Post-double-selection | −0.104 | 0.107 | −0.030 | 0.055 | −0.125 | 0.151 |
| Post-double-selection+ | −0.082 | 0.106 | −0.031 | 0.057 | −0.068 | 0.200 |

**Figura 4 – Print da Tabela de resultados do exemplo empírico de BCH14 (p. 638)**

BCH14 ainda discutem o que chamam de "tradeoff" entre controlar muitas ou poucas variáveis em uma especificação econométrica de problema com dados e modelo HD:

> *"Estamos diante de um tradeoff entre o controle de muito poucas variáveis, o que pode nos deixar imaginando se incluímos controles suficientes para [manter] a exogeneidade da variável de tratamento e o controle de tantas variáveis que seríamos incapazes de aprender de uma maneira mecânica sobre o efeito do tratamento. Os métodos de seleção de variáveis desenvolvidos neste artigo oferecem uma solução para essa tensão. A estrutura esparsa assumida mantém que há um conjunto pequeno o suficiente de variáveis que alguém poderia aprender sobre o tratamento, mas adiciona flexibilidade substancial ao caso usual, onde um pesquisador considera apenas algumas variáveis de controle, permitindo que este conjunto seja encontrado pelos dados de um grande conjunto de controles. Assim, a abordagem deve complementar a análise de especificação cuidadosa usual, fornecendo ao pesquisador uma maneira eficiente e orientada por dados para procurar um pequeno conjunto de fatores de confusão influentes entre um amplo conjunto de variáveis de confusão em potencial escolhido de maneira sensata."*

> *"No exemplo do aborto, usamos o estimador pós-seleção dupla (...) para cada uma de nossas variáveis dependentes. Para o crime violento, oito variáveis são selecionadas na equação do aborto, e nenhuma variável é selecionada na equação do crime. Para o crime contra a propriedade, nove variáveis são selecionadas na equação do aborto, e três são selecionadas na equação do crime. Para o homicídio, nove variáveis são selecionadas na equação do aborto e nenhuma foi selecionada na equação do crime."* (BCH14, pp. 638-639, com grifos próprios, tradução livre e com remoção das notas de rodapé).

Os resultados dos autores utilizando PDS-LASSO são também imprecisos, mas eles promovem o seguinte balanço final:

> *"Acreditamos que o exemplo nesta seção ilustra como podemos usar técnicas modernas de seleção de variáveis para complementar a análise de causalidade em economia. No exemplo do aborto, podemos pesquisar entre um grande conjunto de controles e transformações de variáveis ao tentar estimar o efeito do aborto sobre o crime. A consideração de um grande conjunto de controles torna mais plausível o pressuposto subjacente de exogeneidade da taxa de aborto condicional a variáveis observáveis, enquanto os métodos que desenvolvemos nos permitem produzir um modelo final de dimensão administrável. Curiosamente, vemos que se tiraria conclusões bastante diferentes das estimativas obtidas usando a seleção formal de variáveis. Olhando para as variáveis selecionadas, também podemos ver que esta mudança na interpretação está sendo conduzida pelo método de seleção de variáveis selecionando diferentes variáveis (...) do que normalmente são considerados."* (BCH14, pp. 640, com grifos próprios e tradução livre).

### VIII.3. PÓS-REGULARIZAÇÃO POR LASSO (PR-LASSO)

A metodologia de "Pós-Regularização" – *Post-Regularization*, aqui denominada de PR-LASSO, mas denominada de "Metodologia CHS" pelo pacote *lassopack* – é o conjunto de procedimentos descritos em CHS15. A PR-LASSO é bastante relacionada com a metodologia PDS-LASSO, dado que a dupla seleção por LASSO é feita em ambas as técnicas. Entretanto, ao contrário do PDS-LASSO, o PR-LASSO não utiliza os controles selecionados pelo LASSO no passo final, de regressão OLS pós-regularização. Há um procedimento intermediário, no qual esses controles selecionados são usados para a previsão dentro da amostra, visando construir versões ortogonalizadas da variável dependente e das variáveis causais exógenas de interesse. Essas versões ortogonalizadas são calculadas tendo por base os valores e os coeficientes estimados dos controles selecionados. Há duas versões do estimador:

- a primeira usa os coeficientes estimados da regressão LASSO, construindo "Variáveis ortogonalizadas pelo LASSO" (*LASSO-orthogonalized variables*). As variáveis ortogonalizadas são calculadas a partir da diferença entre a variável original e a previsão a partir dos coeficientes estimados pelo LASSO.
- a segunda usa os coeficientes estimados em uma regressão pós-LASSO, ou seja, para cada um dos passos de seleção de modelo, aplica o estimador OLS para obter coeficientes, construindo "Variáveis ortogonalizadas pelo pós-LASSO" (*post-LASSO-orthogonalized variables*). As variáveis ortogonalizadas são calculadas a partir da diferença entre a variável original e a previsão a partir dos coeficientes estimados pelo OLS pós-LASSO.

De acordo com CHS15, citando Belloni & Chenozukhov (2013), os resultados teóricos da abordagem independem se usamos o estimador LASSO ou o estimador pós-LASSO para o procedimento por eles sugeridos (CHS15, p. 4). Demonstram os autores que o OLS pós-LASSO (ou estimador pós-LASSO) desempenha tão bem quanto o estimador LASSO sob suposições adicionais moderadas. Ou seja, sugerem que qualquer uma das versões do estimador pode ser usada na abordagem de dupla-seleção.

O procedimento CHS15 é importante como uma possibilidade de checagem da robustez dos resultados apresentados pelo PDS-LASSO. Mais do que isso, o procedimento, por reduzir a dimensão da análise para focar apenas no regressando e na(s) variável (variáveis) de interesse, o que tem influenciado artigos mais recentes dos autores que utilizam outros algoritmos em um procedimento denominado de "*Double Machine Learning*" – vide Chernozhukov et al (2018). Por fim, cumpre dizer que o procedimento CHS15 também pode ser aplicado no contexto de regularização por instrumentos (CHS15, p. 487, por exemplo, Algoritmo 1), conforme veremos a seguir com o IV-LASSO.

## VIII.4. Pós-Dupla Seleção com Regularização de Instrumentos (IV-LASSO)

O trabalho BCCH12 busca oferecer uma metodologia que seja adequada a problemas de alta dimensão com relação à definição de variáveis instrumentais – *instrumental variables*, IV – para a correção da endogenia de uma ou mais variáveis de baixa dimensão do modelo. Em suma, temos a situação de que, para endereçar o problema de correlação entre alguns regressores e o termo de erro da regressão, tem-se em mãos um um conjunto muito alto de possíveis instrumentos a serem usados, sendo que um subconjunto considerável deles pode não ser relevante – i.e., coeficientes nulos no primeiro estágio dos estimadores IV. Tem-se, assim, uma outra vertente da maldição da dimensionalidade, em que, nesse caso, adota-se o princípio da esparsidade na identificação do modelo e não no controle de fatores não observáveis da regressão.

BCCH12 iniciam sua análise reafirmando a importância das técnicas de variáveis instrumentais (IV), que são largamente utilizadas em pesquisa econômica aplicada. "*Embora esses métodos forneçam uma ferramenta útil para identificar efeitos estruturais de interesse, sua aplicação geralmente resulta em inferência imprecisa.*" (p. 2370). De fato, o problema mais típico que emerge pós-instrumentação de modelos caracterizados por endogenia é a geração de estimativas com baixa significância estatística.

Explicam os autores que desde estudos como Amemiya (1974), Chamberlain (1987) e Newey (1990), os pesquisadores da área vêm procurando uma forma de melhorar a precisão dos estimadores de variáveis instrumentais, e a forma de equacionar essa questão tem sido

1. tentar alguma aproximação aos instrumentos ótimos, no sentido de melhorar a eficiência dos estimadores; ou
2. usar muitos instrumentos.

No caso de instrumentos ótimos, a abordagem geralmente é feita de forma não paramétrica e, portanto, implicitamente faz uso de muitos instrumentos construídos, como polinômios. BCCH12 comentam que a melhoria prometida na eficiência é atraente, mas os estimadores IV baseados em muitos instrumentos podem ter propriedades ruins. Sendo assim, os autores buscam contribuir com a literatura de estimação IV com muitos instrumentos (caso 2).

A proposta metodológica de BCCH12 é a estimação IV com muitos instrumentos, com uso de modelos LASSO e pós-LASSO para estimar a regressão de primeiro estágio, ou seja, a etapa em que as variáveis regressoras endógenas são rodadas contra os instrumentos – e os demais regressores exógenos e controles. A metodologia dos autores prevê que o estimador LASSO resultante seleciona instrumentos e estima os coeficientes de regressão do primeiro estágio por meio de seu procedimento de encolhimento. O estimador pós-LASSO de BCCH12 procede de maneira similar aos modelos pós-dupla seleção analisados anteriormente, ou seja, descarta as estimativas de coeficiente LASSO para apenas usar a informação de quais instrumentos foram selecionados por ele – o que eles denominam de "conjunto de instrumentos dependente dos dados" – *data-dependent set of instruments*. Esse procedimento é denominado de "IV-LASSO" no descritivo das rotinas do *lassopack.*

Como nos estimadores pós-dupla seleção, o procedimento de BCCH12 realiza a reestimação do modelo IV, dessa vez usando esse conjunto selecionado de variáveis instrumentais e, portanto, reestima as regressões de primeiro estágio, dessa vez usando o OLS, como em um procedimento em Mínimos Quadrados em Dois Estágios (*Two-Stage Least Squares*, 2SLS) tradicional. A diferença desse procedimento para o tradicional procedimento 2SLS em situação de dados HD é que o procedimento BCCH12 efetua a pré-seleção dos instrumentos por meio de regularização por LASSO. Discutem os autores que o procedimento pós-LASSO é realizado para aliviar o viés de encolhimento de LASSO – mesma justificativa do PDS-LASSO, só que nesse caso, aplicada à(s) regressão (regressões) de primeiro estágio. Também enfatizam que "*O uso de métodos baseados em LASSO para formar previsões de primeiro estágio na estimativa IV fornece uma abordagem prática para obter os ganhos de eficiência do uso de instrumentos ótimos enquanto atenua os problemas associados a muitos instrumentos*" (BCCH12, p. 2370).

Com relação às questões de dimensionalidade, a estimação por IV-LASSO assume esparsidade aproximada para o primeiro estágio do estimador – ou seja, existe um pequeno conjunto de instrumentos importantes cujas identidades são desconhecidas que se aproximam bem da esperança condicional das variáveis endógenas dados os IVs. Similarmente ao PDS-LASSO, BCCH12 elencam as vantagens do procedimento IV-LASSO:

- produzem previsões de primeiro estágio que fornecem boas aproximações para os instrumentos ideais, mesmo quando o número de instrumentos disponíveis é muito maior do que o tamanho da amostra em situação do primeiro estágio ser aproximadamente esparso.

- produz estimadores IV que são "consistente raiz-$n$" (*root-n consistent, ou $\sqrt{n}$-consistent*), e assintoticamente normais;
- atinge eficiência semiparamétrica limitada sob a condição adicional de que os erros estruturais são homoscedásticos.
- flexibilidade ao permitir a seleção de modelo imperfeito e, ao mesmo tempo, não impor condições "beta-min" – restrição da magnitude mínima permitida dos coeficientes em regressores considerados relevantes ou "indispensáveis".
- estimador de variância assintótica consistente, que generalizam o procedimento IV da literatura anterior, como Newey (1990) com base na aproximação de séries convencionais dos instrumentos ótimos.
- fornece conjuntos de inferência e intervalos de confiança para o estimador IV do segundo estágio com base nas estimativas LASSO ou pós-LASSO das previsões do primeiro estágio.
- as propriedades são válidas na presença de heteroscedasticidade e, portanto, fornecem um complemento útil para as abordagens existentes na literatura de muitos instrumentos, que muitas vezes dependem de homoscedasticidade e que podem ser inconsistentes na presença de heteroscedasticidade;
- os procedimentos de seleção de instrumentos complementam os métodos existentes/tradicionais que se destinam a ser robustos para muitos instrumentos, mas não são uma solução universal para este problema.
- Ao contrário dos métodos tradicionais de IV, os procedimentos de seleção de instrumentos não requerem que a identidade dessas variáveis "importantes" ou "indispensáveis" seja conhecida *a priori*, pois a identidade desses instrumentos será estimada a partir dos dados.

Como limitação da abordagem proposta, BCCH12 enfatizam que a flexibilidade da modelagem de instrumentos HD vem com o custo de que a seleção de instrumentos tende a não funcionar bem caso o primeiro estágio não seja aproximadamente esparso. Quando o princípio da esparsidade aproximada dos instrumentos não for válido, os procedimentos de seleção de IVs podem ficar instáveis, seja selecionando muito poucos ou nenhum instrumento, seja selecionando instrumentos demais. Dois cenários onde essa falha é provável de ocorrer são o caso dos chamados "instrumentos fracos" e dos "muitos instrumentos fracos". Instrumentos são ditos como "fracos" quando são pouco relacionados às variáveis endógenas que devem instrumentar. BCCH12 estudam duas modificações em seu procedimento básico de estimação IV-LASSO com o objetivo de aliviar esse problema:

- um procedimento de teste de "*sup-score*" – um teste de hipóteses que usa uma estatística equivalente à de razão de verossimilhança (Song, Kosorok, & Fine, 2009) – que está relacionado aos clássicos estudos de instrumentos fracos de Anderson e Rubin (1949) e Staiger e Stock (1997), sendo mais adequado para casos com muitos instrumentos; e
- um estimador IV de amostra dividida que combina a seleção de instrumentos via Lasso com o método de divisão de amostra de Angrist e Krueger (1995).

BCCH12, entretanto, sugerem que pesquisas adicionais devem ser realizadas nesse tema de instrumentos fracos, seja em problemas HD ou não.

Anos após a publicação de BCCH12, o trabalho de CHS15 – o mesmo que introduziu o PR-LASSO – expandiu e generalizou a metodologia de estimação por IV-LASSO, para considerar o caso da existência de alta dimensão tanto nas IVs – usado em BCCH12 – quanto nos controles da regressão – usado em BCH14. Utilizaremos esse arcabouço na exposição a seguir.

Considere o seguinte modelo linear IV contendo muitos controles e muitos instrumentos (CHS15):

$$y_i = \alpha d_i + x_i'\beta + \varepsilon_i \tag{19}$$

$$d_i = x_i'\gamma + z_i'\delta + u_i \tag{20}$$

onde, como visto anteriormente em (14) e (15), $y_i$ é a variável dependente, $i$ representa a $i$-ésima observação, $d_i$ é uma variável de interesse, $x_i = (x_1, \ldots, x_{p_x})$ é um vetor $p_x$-dimensional que engloba todos os controles de alta dimensão, $\beta$ e $\gamma$ são vetores de parâmetros, e $\varepsilon_i$ é o termo de erro aleatório. A diferença está na equação (16), com o termo $z_i'\delta$. $z_i$ é o vetor de alta dimensão de variáveis instrumentais, $z_i = (z_1, ..., z_{p_z})$..

O algoritmo de estimação usado pela metodologia do IV-LASSO é similar ao do PDS-LASSO, e também utiliza três passos, sendo os dois primeiros de regularização/seleção e o último, uma regressão final. As duas diferenças estão nos passos II e III, com, respectivamente, a inclusão das variáveis instrumentais adicionalmente aos controles, e a regressão por método IV em vez de OLS. O algoritmo é o seguinte (BCCH12):

- **Passo I (Primeira Seleção de Modelos)**: estimar uma regressão LASSO contendo $y_i$ como a variável dependente e as variáveis de controle $x_i$ como regressores penalizados;
- **Passo II (Segunda Seleção de Modelos)**: estimar uma regressão LASSO contendo $d_i$ como a variável dependente, e tendo as variáveis de controle $x_i$ e as variáveis instrumentais $z_i$ como regressores penalizados;
- **Passo III (Estimação Final)**: rodar uma regressão linear do tipo IV (2SLS) contendo $y_i$ como a variável dependente, e, como regressores, $d_i$ e também a união do conjunto de variáveis $x_i$ selecionados nos Passos I e II, e contendo as variáveis instrumentais $z_i$ selecionadas no passo II para endereçar a endogeneidade de regressores no modelo.

## IX. O PACOTE DE ROTINAS LASSOPACK

Ahrens, Hansen & Schaffer (2019, 2020) desenvolveram o pacote de rotinas denominado de "*lassopack*", disponível no link statalasso.github.io/docs/lassopack. As rotinas desenvolvidas permitem rodar modelos de regressão do tipo *Post-Double Selection* (PDS-LASSO e IV-LASSO) e *Post-Regularization* (PR-LASSO, denominada por eles de "metodologia CHS").

Às custas de um abuso de siglas, nos referiremos ao artigo publicado pelos autores no Stata Journal como "AHS20", e aos comandos propriamente ditos, o seu *help* e site do *lassopack*, como "AHS19". AHS19 enfatizam que não se tratam de comandos oficiais do Stata, mas contribuições gratuitas para a comunidade de pesquisa.

### IX.1. SINTAXE

Existem duas rotinas principais, "pdslasso" e "ivlasso", que podem ser baixadas digitando-se **search lassopack, all**. Para instalar diretamente, consultar statalasso.github.io/installation. A rotina ivlasso é mais geral, dado que permite rodar os mesmos modelos da rotina pdslasso, além de rodar os seus modelos IV, que são específicos. A sintaxe dos comandos é a seguinte:

```
pdslasso depvar regressors (hd_controls) [weight] [if exp] [in range] [ ,
 partial(varlist) pnotpen(varlist) aset(varlist) post(method) robust
 cluster(var) fe noftools rlasso[(name)] sqrt noisily loptions(options)
 olsoptions(options) noconstant ]
```

```
ivlasso depvar regressors [(hd_controls)] (endog=instruments) [if exp] [in
 range] [ , partial(varlist) pnotpen(varlist) aset(varlist) post(method)
 robust    cluster(var)    fe    noftools    rlasso[(name)]    sqrt    noisily
 loptions(options) ivoptions(options) first idstats sscset ssgamma(real)
 ssgridmin(real)     ssgridmax(real)     ssgridpoints(integer     100)
 ssgridmat(name) noconstant ]
```

Note que tudo o que estiver entre colchetes se refere a configurações opcionais. Por exemplo "depvar" (variável dependente) e "regressors" (variáveis explicativas), são de inserção obrigatória, enquanto "pnotpen(varlist)" (variáveis explicativas não penalizadas), são opcionais. Se uma configuração – obrigatória ou não – estiver entre parênteses ou for seguido de parênteses, isso significa que ela necessitará desse elemento para ser rodada; por exemplo, no caso de "pnotpen(varlist)", a configuração "pnotpen" requer uma lista de variáveis entre parênteses "(varlist)".

Para fins de simplicidade da exposição, consideremos apenas a rotina ivlasso, que é mais geral – ou seja, todos os comandos da rotina pdslasso podem ser executados também utilizando aquela rotina. Assim, que o comando ivlasso reproduz exatamente os resultados do pdslasso e ainda roda os seus próprios resultados.

Utilizaremos como exemplo abstrato um problema de alta dimensão extremo – número de controles e de variáveis instrumentais bem maior do que o número de observações –, em uma amostra contendo 1000 observações, onde o regressando é $y$, os regressores de baixa dimensão são compostos por $x_1, x_2, ..., x_{10}$, os regressores (controles) de alta dimensão são compostos por $c_1, c_2, ..., c_{5000}$, e as variáveis instrumentais são $z_1, z_1,..., z_{2000}$. O exemplo é generalizável para o caso HD não tão extremo, em que o número de controles e variáveis instrumentais pode ser absolutamente alto, mas não necessariamente relativo ao tamanho da amostra.

Considere os casos de destaque a seguir.

### IX.1.1. Modelo HD

Regressão OLS tradicional, com uso de todos os controles HD:

```
regress y x1-x10 c1-c5000
```

Nesse caso, ao requerer o comando "regress", obtemos o resultado de um estimador OLS, com y rodado contra as variáveis de baixa e alta dimensão, $x_1$ a $x_{10}$ e $c_1$ a $c_{5000}$. Note que o estimador OLS é irrestrito com relação aos parâmetros possíveis, não efetuando penalizações. Como não há graus de liberdade suficientes (5010 + 1 parâmetros a estimar, para 1000 observações), essa regressão não pode ser estimada. O pesquisador pode impor alguma simplificação na forma de seleção de subconjuntos (*subset selection*) de $c$, para viabilizar a estimação – incorrendo nos problemas conhecidos desse tipo de abordagem, com *p-hacking* e *overfitting*.

### IX.1.2. Modelo HDS simples

Regressão PDS-LASSO, com uso de controles "Esparsos de Alta Dimensão" (*High-Dimension Sparse*, HDS) e todas as variáveis de baixa dimensão sendo "variáveis focais":

```
ivlasso y x1-x10 (c1-c5000)
```

Nesse caso, temos a regressão PDS-LASSO, com y rodado contra dois conjuntos de regressores, $x_1$ a $x_{10}$ ("baixa dimensão", LD), e $c_1$ a $c_{5000}$ ("alta dimensão", HD). Note que, por estarem fora dos parênteses, nenhuma das variáveis $x$ é penalizada – ou seja, todas as variáveis são configuradas como "focais" pelo pesquisador. Lembrando da discussão da Seção 7, temos que variáveis focais – ou "de geração" – são aquelas sempre incluídas no modelo por haver alguma crença do pesquisador que elas pertencem ao PGD. Note que, dessas variáveis $x$, talvez apenas um subconjunto delas ou mesmo apenas uma, seja "variável de interesse". Os regressores $c_1$ a $c_{5000}$ são aproximações a "fatores de confusão" (*confounding factors*), estimados na forma de "controles" ou "variáveis de incômodo" (*nuisance variables*). Por estarem entre parênteses, esses regressores são penalizados pelo comando ivlasso, sendo que a solução final conterá apenas um subconjunto não nulo deles – solução esparsa permitida pelo LASSO. Ou seja, muitos dos controles HD, $c_1$ a $c_{5000}$, serão inativados pela regularização.

### IX.1.3. Modelo HDS conservador

Nesse caso, temos uma regressão PDS-LASSO, com uso de controles "Esparsos de Alta Dimensão" (*High-Dimension Sparse*, HDS) e com penalização de variáveis de baixa dimensão:

```
ivlasso y x1 (x2-x10 c1-c5000)
```

Nesse caso, temos a regressão PDS-LASSO, com y rodado contra três conjuntos de regressores: $x_1$, $x_2$ a $x_{10}$, e $c_1$ a $c_{5000}$. Por estar fora dos parênteses, $x_1$ é não penalizada, sendo a variável focal do estudo, conforme configuração do pesquisador. Note que a rotina/modelo PDS requer no mínimo uma dessas variáveis – ou seja, pelo menos um regressor deverá está imune ao procedimento de regularização/inativação. Consistente com Danilov & Magnus (2004), denominados as demais variáveis $x$ de "variáveis auxiliares". Note também que não necessariamente $x_1$ é a "variável de interesse" do estudo. Pode ser que uma ou mais variáveis de $x_2$ a $x_{10}$ seja "de interesse", e que o objetivo do pesquisador é justamente passar essas variáveis no crivo do LASSO. Ao expor uma variável de interesse ao processo de penalização, o pesquisador estará adotando um procedimento precavido, assumindo que a variável pode não pertencer ao PGD e que, somente possuirá evidências contrárias disso caso a variável resista à regularização e, na pós-seleção, mantenha-se estatisticamente significante. Por esse motivo, esse modelo pode ser denominado de "HDS conservador".

Caso o pesquisador tenha outras variáveis focais dentre os regressores LD – por exemplo, $x_2$ e $x_3$, além de $x_1$ –, basta retirá-las dos parênteses. O comando fica, assim, da seguinte forma:

```
ivlasso y x1-x3 (x4-x10 c1-c5000)
```

Nesse caso, a penalização efetuada pela regressão PDS-LASSO, não irá ser aplicada a $x_1$, $x_2$ e $x_3$, mas aplicada às variáveis auxiliares $x_3$ a $x_{10}$, e aos controles HD esparsos $c_1$ a $c_{5000}$. Nesse caso, como no anterior, a estimação como "parâmetro de incômodo" aplica-se apenas a $c_1$ a $c_{5000}$, sendo que os parâmetros das variáveis auxiliares não recebem essa interpretação, mesmo que sofram de possível inativação pelo LASSO. Apesar de estarem sujeitas a regularização, as variáveis auxiliares são de baixa dimensão, e a sua inativação está mais relacionada ao pertencimento ao PGD do que ao princípio da esparsidade que é aplicável aos controles HD $c_1$ a $c_{5000}$.

## IX.1.4. Modelo IV-HDS

Regressão IV-LASSO, com uso de controles HDS, e uma variável endógena instrumentada por variáveis instrumentais também HDS:

```
ivlasso y (x1 = z1-z2000) x2-x10 (c1-c5000)
```

Nesse caso, regressão IV-LASSO, com y rodado contra $x_1 - x_{10}$ e $c_1$ a $c_{5000}$. A diferença nesse caso está no novo termo entre parênteses "(x1 = z1-z2000)", onde que $x_1$ é uma variável endógena – ou seja, correlacionada com o erro da regressão –, e $z_1$ a $z_{2000}$ são variáveis instrumentais esparsas penalizadas. Continuamos tendo $x_2$ a $x_{10}$ com variáveis LD (focais) e $c_1$ a $c_{5000}$ são controles HDS estimados como parâmetros de incômodo. Na rotina ivlasso, toda e qualquer variável endógena será sempre ativa, ou seja, as endógenas nunca são penalizadas pelo LASSO. Por utilizar variáveis instrumentais e controles em uma situação HD, denominamos esse procedimento de "IV-HDS". Note que fica a cargo do pesquisador definir quais outras variáveis LD de seu modelo são penalizadas. Abaixo temos outras variantes do modelo IV-HDS, dentre as possíveis.

- Uma variável endógena (não penalizada), 9 variáveis auxiliares (penalizadas), 5000 controles HDS e 2000 variáveis instrumentais HDS:

```
ivlasso y (x1 = z1-z2000) (x2-x10 c1-c5000)
```

- Duas variáveis endógenas (não penalizadas), 8 variáveis auxiliares (penalizadas), 5000 controles HDS e 2000 variáveis instrumentais HDS:

```
ivlasso y (x1 x2 = z1-z2000) (x3-x10 c1-c5000)
```

- Duas variáveis endógenas (não penalizadas), duas variáveis focais (não penalizadas), 6 variáveis auxiliares (penalizadas), 5000 controles HDS e 2000 variáveis instrumentais HDS:

```
ivlasso y (x1 x2 = z1-z2000) x3 x4 (x5-x10 c1-c5000)
```

- Duas variáveis endógenas (não penalizadas), duas variáveis focais (não penalizadas), 6 variáveis auxiliares (penalizadas), 5000 controles HDS e 1998 variáveis instrumentais HDS e duas variáveis instrumentais não penalizadas:

```
ivlasso y (x1 x2 = z3-z2000) x3 x4 (x5-x10 c1-c5000), pnotpen(z1-z2)
```

- Duas variáveis endógenas (não penalizadas), duas variáveis focais (não penalizadas), 6 variáveis auxiliares (penalizadas), 5000 controles HDS e 1998 variáveis instrumentais HDS, duas variáveis instrumentais não penalizadas e 3 regressores adicionais, que comporão a regressão final do PDS – conjunto de "melhoria" (*amelioration set*) –, mas que não passaram pelas etapas de LASSO:

```
ivlasso y (x1 x2 = z3-z2000) x3 x4 (x5-x10 c1-c5000),
    pnotpen(z1-z2) aset(x11 x12)
```

Note que o conjunto de "melhoria" é uma opção de especificação ao pesquisador, que deverá justificar este procedimento – e os demais procedimentos adotados – em sua exposição da modelagem.

### IX.2. Penalidades específicas

A expressão para o LASSO adotada por Ahrens, Hansen & Schaffer (2020) é a seguinte:

$$\hat{\beta}_{LASSO}(\lambda) = \arg\min_{\beta}\left\{\frac{1}{N}\sum_{i=1}^{N}\left(y_i - \sum_{j=1}^{p}\beta_j x_{j,i}\right)^2 + \frac{\lambda}{N}\sum_{j=1}^{p}\psi_j|\beta_j|\right\}$$

onde a divisão por $N$ se dá pelo fato de considerarem a minimização do erro quadrático médio. A configuração dos autores segue o modelo mais geral de BCCH12 (p. 2379) e BCH14 (p. 615), sendo mais flexível que o arcabouço geral do LASSO por permitir a existência de cargas de penalidade que são específicas do regressor – "*predictor-specific penalty loadings*", denotadas com $\psi_j$, onde $j$ é o $j$-ésimo regressor. Dessa forma, $\lambda$ controla o nível de penalidade geral sobre a magnitude dos coeficientes, enquanto $\psi_j$ controla o nível de penalidade específico de cada um deles.

Explicam os Ahrens, Hansen & Schaffer (2020, p. 180) que o procedimento de pré-padronização das variáveis – ou seja, transformá-las em variáveis com média 0 e variância 1 – é equivalente ao procedimento de escolha de cargas de penalidade específicas para levar em consideração a variância desigual específica de cada regressor. Outra característica das rotinas do *lassopack* é que estas permitem a estimação de cargas de penalidade "robustas" ou "cluster-robustas", de forma a levar em consideração problemas de heteroscedasticidade na regressão. Abaixo seguem exemplos.

- PDS-LASSO com estimativas HAC (*heteroskedastic-robust*):

```
ivlasso y x1 (x2-x10 c1-c5000), rob
```

- PDS-LASSO com estimativas *cluster-robust*:

```
ivlasso y x1 (x2-x10 c1-c5000), cluster(w1)
```

onde $w_1$ é uma variável de agrupamento (*cluster*) presente na base de dados.

### IX.3. Pós-seleção

Consistente com a metodologia de "pós-seleção", discorrem Ahrens, Hansen & Schaffer (2020) que os métodos de regressão penalizada induzem um viés de atenuação – isto é, o enviesamento da inclinação da regressão para zero (a subestimação de seu valor absoluto), também chamado de "diluição da regressão". Esse viés pode ser aliviado rodando-se um estimador OLS pós-estimação, que aplica OLS às variáveis selecionadas pelo método de seleção de variável de primeiro estágio. Assim, temos:

$$\hat{\beta}_{pós} = \arg\min_{\beta}\left\{\frac{1}{N}\sum_{i=1}^{N}\left(y_i - \sum_{j=1}^{p}\beta_j x_{j,i}\right)^2\right\}$$

$$\text{sujeito a } \beta_j = 0 \text{ se } \tilde{\beta}_j = 0,$$

onde $\tilde{\beta}_j$ é um estimador esparso de primeira etapa, nesse caso, o LASSO. Dessa forma, a pós-estimativa OLS trata o estimador da primeira etapa como uma técnica genuína de seleção de modelo. Utilizando um modelo LASSO com regularização orientada por teoria, Belloni e Chernozhukov (2013) mostraram que a pós-estimativa OLS, também conhecida como "pós-LASSO", atinge as mesmas taxas de convergência que o LASSO e pode superar o LASSO em situações onde a seleção de modelo consistente é viável. Vide discussão em BCCH12.

## IX.4. Dados em painel

Nas rotinas do *lassopack* (pdslasso e ivlasso) é possível efetuar a estimação com procedimento de estimador de efeitos fixos, quando a estrutura de dados for em painel. Essa opção de efeitos fixos é equivalente a especificar dummies específicos de painel não penalizado, ou seja, atribuir a cada indivíduo do painel uma variável binária *one-hot encoding* e que não sejam sujeitas a regularização/inativação pelo LASSO. Entretanto, o procedimento das rotinas é computacionalmente mais rápido e mais preciso do que o uso de *dummies*, dado que utiliza o tradicional passo de transformação "*within*" do estimador de efeitos fixos, onde se extrai a média interna de cada variável – nesse caso, não há estimativa de constante e nem dos parâmetros de efeitos fixos. A variável de painel usada pela opção fe é a variável de painel previamente definida por *xtset* – comando do Stata para esse tipo de designação.

- PDS-LASSO com regressão de efeitos fixos para dados em painel:

```
ivlasso y x1 (x2-x10 c1-c5000), fe
```

## IX.5. Escolha dos parâmetros de ajuste

Um ponto fundamental de modelagem de qualquer procedimento de regularização diz respeito à escolha do(s) hiperparâmetro(s) de penalização de variáveis. Ou seja, a seleção dos parâmetros de ajuste de uma maneira otimizada torna-se questão vital de confiabilidade das estimativas produzidas pelas rotinas. Dizem Ahrens, Hansen & Schaffer (2020): "*Qual método é mais apropriado depende dos objetivos e da configuração efetuada, particularmente do objetivo da análise (previsão ou identificação do modelo), restrições computacionais, e se e como o pressuposto de resíduos independente e identicamente distribuídos (i.i.d.) é violado.*" (p. 184, tradução livre)

Consistente com as discussões em BCCH12, BCH14 e CHS15 – e outros trabalhos na área de pós-dupla seleção por LASSO –, Ahrens, Hansen & Schaffer (2020) discorrem que a *lassopack* oferece três abordagens para selecionar o valor "ótimo" de $\lambda$ – e também $\alpha$, no caso de um modelo de Rede Elástica. As abordagens utilizadas no *lassopack*, denominadas de escolha dos hiperparâmetros "conduzidas pelos dados" (*data-driven approach*), são as seguintes:

- <u>Abordagem do critério de informação</u>: critérios como o de Akaike (*Akaike Information Criterion, AIC*), e o Bayesiano (*Bayesian Information Criterion*, BIC), e equivalentes estendidos – por exemplo, o EBIC –, calculados a partir da rotina <u>lasso2</u>.
- <u>Abordagem da validação cruzada ("*cross validation*", CV)</u>: objetivo de otimizar o desempenho de previsão fora da amostra do modelo. As validações cruzadas dos tipos "*K-fold*" e "*rolling*" – tanto para dados de painel quanto para série temporal – são implementadas na rotina <u>cvlasso</u>. O problema da abordagem CV é que ela pode ser bastante demandante em termos computacionais. Discutem os autores que "*Se o objetivo da análise for identificar o modelo verdadeiro, o BIC e o EBIC fornecem uma escolha melhor do que o K-fold CV, porque existem condições sob hipóteses fortes, mas bem compreendidas, nas quais o BIC e o EBIC são consistentes na seleção de modelos.*"
- <u>Abordagem "rigorosa" (*rigourous*), ou orientada pela teoria (*theory-driven*)</u>: Consistente com Chernozhukov, Hansen e Spindler (2016), utiliza-se o termo "rigoroso" para enfatizar que o arcabouço é baseado em teoria, na medida em que fornece parâmetros de penalização que garantem a taxa ótima de convergência para previsão e estimativa de parâmetros. A abordagem também produz modelos cujo tamanho é da mesma ordem que o modelo verdadeiro Ahrens, Hansen & Schaffer (2020, p. 190). Sua origem está em BCCH12, que parte do pressuposto de que a variância do termo de erro é desconhecida e propõem uma metodologia denominada de "LASSO rigoroso" (ou "LASSO viável" – *rigorous* LASSO ou *feasible* LASSO), que se baseia em um algoritmo iterativo para estimar a penalização ótima, sendo também válido na presença de erros não gaussianos e heteroscedásticos. Em um outro estudo, Belloni et al. (2016) realizam uma extensão desse modelo para uma estimação com dados em painel. A penalização rigorosa é a forma utilizada de se obter o parâmetro de ajuste nas rotinas *pdslasso* e *ivlasso* do *lassopack*. A sub-rotina utilizada para implementar a penalização rigorosa é o comando <u>*rlasso*</u>.

### IX.6. OUTROS MODELOS DE REGULARIZAÇÃO DO LASSOPACK

Conforme discutem Ahrens, Hansen & Schaffer (2020), as rotinas do *lassopack* (pdslasso e ivlasso) permitem rodar diversos outros modelos de regularização. É possível rodar regressão RIGDE, Elastic Net (Rede Elástica), Adaptive LASSO, Square root LASSO, Rigorous LASSO, Rigorous Square Root LASSO, com ou sem estimador de efeitos fixos, variáveis instrumentais e controles de heteroscedasticidade.

## X. APLICAÇÕES RECENTES NA LITERATURA PUBLICADA E NÃO PUBLICADA

Apesar de ainda estar apenas no início de desenvolvimento, a literatura empírica que utiliza modelos de Pós-Dupla Seleção por LASSO ("PDS-LASSO"), Pós-Regularização ("PR-LASSO", mais conhecido como "metodologia CHS") e Pós-Regularização de Modelos de Variáveis Instrumentais com LASSO ("IV-LASSO"), mostra sinais de que em breve deve ser bastante ampla e variada. Em uma pesquisa de referências em outubro de 2020, foram encontrados pouco mais de duas dezenas de trabalhos que utilizam a técnica, sendo que a maioria ainda não foi publicada. Trata-se de uma área emergente e possivelmente vibrante. Abaixo listamos alguns desses trabalhos.

- Azoulay, Greenblatt & Heggeness (2019) investigam, com uso de PDS-LASSO, o efeito de uma exposição relativamente curta e intensa, à pesquisa de fronteira nas trajetórias de carreira de inovadores potenciais? Os autores pertencem ao MIT Sloan School of Management, e U.S. Census Bureau. O estudo é um documento de trabalho do National Bureau of Economic Research - NBER, nos Estados Unidos.
- Barrera-Osorio, Gertler, Nakajima & Patrinos (2020) investigam com uso de PDS-LASSO, a promoção de envolvimento dos pais em escolas mexicanas. Os autores pertencem à Vanderbilt University, University of California/Berkeley, Harvard University e World Bank.
- Berggren & Nilsson (2020) examinam como a variação no anti-semitismo entre países pode ser explicada pelo grau de liberdade econômica. Os autores pertencem à University of Economics, República Tcheca, e Lund University, Sweden.
- Blevins & Kawata (2019) examinam o impacto da orfandade nos resultados de aprendizagem entre as meninas e
- meninos na África Subsaariana, condicionada à matrícula escolar. Os autores utilizam PDS-LASSO; pertencem à Hiroshima University e University of Tokyo.
- Cullen (2020), da University of Oxford, investiga com uso de PDS-LASSO, o fenômeno da subnotificação de violência por parceiro íntimo na Nigéria e Ruanda.
- Dean & Jayachandran (2019) investigam, com uso de PDS-LASSO, a mudança de atitude da família para promover o emprego feminino na Índia. Os autores pertencem ao Institute on Behavior and Inequality, na Alemanha, e à Northwestern University (EUA).
- Fluchtmann, Glenny, Harmon & Maibom (2020) investigam com PDS-LASSO se homens e mulheres se candidatam aos mesmos empregos, para entendem as diferenças de gênero no mercado de trabalho. Os autores pertencem à OECD, University of Copenhagen e Aarhus University.
- Heß, Jaimovich & Schündeln (2018) estudam, com uso de PDS-LASSO, os efeitos dos projetos de desenvolvimento nas redes econômicas da Gâmbia Rural. Artigo publicado na Review of Economic Studies (Advance access publication) 2020. Os autores pertencem à Goethe University Frankfurt e Universidad de Talca.

- Hill (2020) é uma tese de doutorado do MIT, defendida em maio de 2020. O estudo utiliza PDS-LASSO para investigar as forças econômicas que influenciam a criação de pesquisas acadêmicas e conhecimentos científicos básicos.
- Jinks, Kniesner, Leeth & Sasso (2020) estudam com PDS-LASSO o caso do Texas, que é o único estado que não exige que os empregadores tenham cobertura de seguro de compensação dos trabalhadores. Os autores pertencem a University of Illinois at Chicago, University of Illinois at Chicago, Bentley University, e DePaul University, todas nos Estados Unidos.
- Kiessling & Norris (2020) utilizam PDS-LASSO para estudar os efeitos de longo prazo de colegas na saúde mental de um estudante. Os autores pertencem, respectivamente, ao Max Planck Institute for Research on Collective Goods, da Dinamarca, e University of Strathclyde, no Reino Unido.
- Macchiavello, Menzel, Rabbani & Woodruff (2020) estudam a participação feminina em funções gerenciais no setor de vestuário em Bangladesh. Os autores pertecem à London School of Economics (LSE), CEGGE-EI de Praga, University of Dhaka, de Bangladesh e Universidade de Oxford. O trabalho consta como documento de trabalho do National Bureau of Economic Research - NBER, nos Estados Unidos.
- McKelway (2019), estuda a participação no mercado de trabalho de mulheres na Índia, utilizado PDS-LASSO. O trabalho é fruto de sua tese de doutorado defendida no MIT em 2020.
- Menzel & Woodruff (2019) é um documento de trabalho de autores da CERGE-EI de Praga e da University of Oxford. A aplicação de PDS-LASSO para selecionar controles é na área de diferenças salariais entre gêneros e mobilidade do trabalho.
- Nguyen (2020) é uma tese da University of Chicago, defendida em agosto de 2020, na área de Economia do Trabalho. O autor compara estimativas OLS, LASSO e PDS-LASSO.
- Peng (2019) é um documento de trabalho postado na plataforma ArXiv que investiga efeitos endógenos heterogêneos para identificar líderes e seguidores em redes de pessoas. Apesar do estudo não utilizar a metodologia BCCH14, utiliza um método próprio para selecionar instrumentos.
- Ricci et al (2020) utilizam PDS-LASSO em estudos de medicina, para estudar efeitos de complicação abrangente em cirurgia pancreática. Participaram da pesquisa diversos autores da Universitaria Di Bologna, Itália.

## XI.1. APLICAÇÃO AO TRANSPORTE AÉREO

Um dos estudos recentes aplicados ao transporte aéreo que utiliza métodos de regressão regularizada é o de Oliveira et al. (2021). Os autores empregam o método LASSO para estimar a eficiência no consumo de combustível no transporte aéreo, levando em conta a gestão dinâmica de frotas (Dynamic Fleet Management, DFM). Eles propõem um modelo econométrico que integra dados detalhados de frota, rede de rotas, preços de combustível e variáveis meteorológicas, com o objetivo de identificar os efeitos causais dos sinais de preço sobre decisões de substituição de aeronaves e eficiência operacional. O trabalho parte do reconhecimento de que a eficiência energética das companhias aéreas depende não apenas de fatores tecnológicos, mas também de estratégias operacionais e gerenciais, muitas delas não observáveis diretamente. Por isso, os autores enfatizam o desafio de controlar tais fatores latentes, que afetam simultaneamente o consumo de combustível e a resposta das empresas aos choques de preços.

Os autores desenvolvem uma metodologia econométrica de alta dimensionalidade baseada no uso do LASSO e de sua versão instrumental (IV-LASSO), inspirada em Belloni, Chernozhukov e Hansen (2012, 2014a,b). Essa abordagem é projetada para lidar com milhares de variáveis correlacionadas, selecionando apenas aquelas que explicam de forma informativa as variações não observáveis associadas às operações aéreas. O LASSO atua como um mecanismo de regularização que permite capturar a heterogeneidade operacional entre aeronaves, rotas e companhias sem comprometer a estabilidade do modelo, identificando subconjuntos parcimoniosos de controles relevantes. Isso é particularmente importante diante da elevada correlação entre variáveis técnicas (como configuração da frota, idade média e densidade de assentos) e variáveis de mercado (como carga transportada e distância média).

Os autores também utilizam o IV-LASSO para tratar potenciais problemas de endogeneidade, permitindo a seleção automática dos instrumentos mais relevantes em um conjunto extenso de candidatos. Essa estrutura fornece inferências mais robustas sobre o efeito causal dos preços de combustível, evitando o viés decorrente de instrumentos fracos e reduzindo o sobreajuste. O modelo desenvolvido pelos autores incorpora controles de DFM representados pela participação de cada matrícula de aeronave nos pares de cidades, variável que reflete as capacidades operacionais específicas de cada companhia aérea. No total, 1.272 controles desse tipo foram considerados, dos quais 405 (31,8%) foram mantidos pelo LASSO, representando os efeitos realmente informativos relacionados à gestão dinâmica da frota. Dessa forma, o método atua como um filtro que elimina as influências espúrias desses fatores não observáveis, estabilizando os coeficientes principais e reduzindo o viés por variáveis omitidas.

Em síntese, os autores efetuam uma aplicação que combina procedimentos de regularização e identificação causal para analisar as relações entre preços de combustível, gestão de frota e eficiência operacional. Os autores discutem que a adoção do LASSO e de sua versão com variáveis instrumentais promovem a precisão das estimativas e o poder explicativo do modelo, ao permitir que a especificação incorpore um número elevado de controles correlacionados e mantenha a consistência dos resultados. Assim, a integração entre regularização e inferência causal utilizada por Oliveira et al. (2021) pode permitir capturar heterogeneidades operacionais não observáveis e gerar estimativas provavelmente mais consistentes dos determinantes da eficiência energética no transporte aéreo.

# XI. Considerações finais

O presente trabalho visou apresentar, de uma maneira didática, uma discussão acerca dos Modelos de Pós-Dupla Seleção e Pós-Regularização por LASSO apresentados recentemente em uma sequência de trabalhos por Alexandre Belloni, Christian B. Hansen e Victor Chernozhukov. Apresentou-se também um conjunto de rotinas denominadas de "lassopack" (Ahrens, Hansen & Schaffer, 2019, 2020) e disponíveis para o software Stata para uso de pesquisadores com dados e modelos de alta dimensão (HD). Busca-se disseminar o conhecimento nessa área, propiciando um uso mais intenso da metodologia na literatura econômica nacional.

As limitações do uso do LASSO são bem conhecidas na literatura, e por esse motivo é importante concluir este trabalho com uma breve menção a elas. Em termos formais, o uso do método LASSO em modelos econométricos deve ser entendido principalmente como um instrumento para estimar a parte de incômodo (nuisance) de um problema e não como um procedimento confiável de seleção de variáveis. Isso ocorre porque o LASSO não é consistente em seleção de variáveis, isto é, ele não garante a identificação exata do verdadeiro conjunto de regressoras relevantes à medida que o tamanho amostral cresce. Zhao e Yu (2006) demonstram que essa inconsistência decorre do não atendimento de uma condição estrutural específica do conjunto de regressoras, denominada Irrepresentable Condition. Essa condição estabelece que as variáveis irrelevantes não podem ser excessivamente correlacionadas com as variáveis relevantes do modelo, pois, quando isso ocorre, as primeiras podem ser "representadas" como combinações lineares das segundas. Em outras palavras, se as variáveis fora do modelo verdadeiro forem altamente correlacionadas com as que pertencem a ele, o LASSO tende a confundir sua contribuição, incluindo regressoras espúrias ou excluindo variáveis importantes. Nesses casos, o método permanece útil para predição, mas não para inferência causal. Qualquer tentativa de interpretação ou inferência sobre os parâmetros de alta dimensionalidade selecionados pelo LASSO exigiria que o método fosse perfeito na seleção do modelo, o que não ocorre, exceto sob hipóteses altamente irreais. Em termos práticos, seu papel é reduzir dimensionalidade e controlar variáveis potencialmente correlacionadas com o termo de erro, permitindo estimativas mais robustas dos parâmetros de interesse. Assim, o LASSO deve ser interpretado como uma ferramenta auxiliar de regularização, adequada para selecionar controles em contextos de alta dimensionalidade, mas não para definir as variáveis focais de um modelo cujo objetivo é a inferência estrutural.

# Referências

## Referências de teóricas da metodologia HDS e Seleção por LASSO

REFERÊNCIAS DE APLICAÇÕES DA METODOLOGIA PDS-LASSO